\documentclass[
aip,
jcp,
showpacs,
amsmath,
amssymb,
preprint,
floatfix,
longbibliography,
letterpaper,
lengthcheck,
superscriptaddress
]{revtex4-1}

\usepackage{graphicx}
\usepackage[colorlinks=true]{hyperref}
\usepackage{braket}
\usepackage{accents}

\newcommand{\norm}[1]{\left|\left|#1\right|\right|}
\newcommand{\toop}[1][]{\mathcal{T}_{#1}}
\newcommand{\op}[1]{\hat{#1}}
\newcommand{\abs}[1]{\left| #1 \right|} 
\newcommand\ubar[1]{\underaccent{\bar}{#1}}
\newcommand{\V}[1]{\vec{#1}}
\newcommand{\M}[1]{\boldsymbol{#1}}
\newcommand{\T}[1]{\boldsymbol{\vec{#1}}}
\newcommand{\Z}{\mathcal{Z}}
\newcommand{\Tr}{\mathrm{Tr}}
\newcommand{\tr}{\mathrm{tr}}
\begin{document}

\title{
Many-body Green's function theory for electron-phonon interactions:
the Kadanoff-Baym approach to spectral properties of the Holstein dimer}

\author{Niko \surname{S{\"a}kkinen}}
\affiliation{
Department of Physics, Nanoscience Center,
University of Jyv{\"a}skyl{\"a},
Survontie 9, 40014 Jyv{\"a}skyl{\"a}, Finland}

\author{Yang \surname{Peng}}
\affiliation{Dahlem Center for Complex Quantum Systems and Fachbereich Physik,
Freie Universit{\"a}t Berlin, 14195 Berlin, Germany}
\affiliation{
Fritz-Haber-Institut der Max-Planck-Gesellschaft,
Faradayweg 4-6, 14195 Berlin-Dahlem, Germany}

\author{Heiko \surname{Appel}}
\affiliation{
Fritz-Haber-Institut der Max-Planck-Gesellschaft,
Faradayweg 4-6, 14195 Berlin-Dahlem, Germany}
\affiliation{
Max-Planck-Institut f\"ur Struktur und Dynamik der Materie, 
Luruper Chaussee 149, 22761 Hamburg, Germany}
\affiliation{European Theoretical Spectroscopy Facility (ETSF)}

\author{Robert \surname{van Leeuwen}}
\affiliation{
Department of Physics, Nanoscience Center,
University of Jyv{\"a}skyl{\"a},
Survontie 9, 40014 Jyv{\"a}skyl{\"a}, Finland}
\affiliation{European Theoretical Spectroscopy Facility (ETSF)}


\begin{abstract}
We present a Kadanoff-Baym formalism to study time-dependent phenomena for
systems of interacting electrons and phonons in the framework of many-body
perturbation theory. The formalism takes correctly into account effects of the 
initial preparation of an equilibrium state, and allows for an explicit
time-dependence of both the electronic and phononic degrees of freedom. The
method is applied to investigate the charge neutral and non-neutral excitation 
spectra of a homogeneous, two-site, two-electron Holstein model. This is an 
extension of a previous study of the ground state properties in the Hartree (H), 
partially self-consistent Born (Gd) and fully self-consistent Born (GD) 
approximations published in Ref.~\onlinecite{saekkinen-2014a}. We show that 
choosing a homogeneous ground state solution leads to unstable dynamics for a 
sufficiently strong interaction, and that allowing a symmetry-broken state
prevents this. The instability is caused by the bifurcation of the ground state
and understood physically to be connected with the bipolaronic crossover of the 
exact system. This mean-field instability persists in the partially
self-consistent Born approximation but is not found for the fully
self-consistent Born approximation. By understanding the stability properties,
we are able to study the linear response regime by calculating the
density-density response function by time-propagation. This functions amounts
to a solution of the Bethe-Salpeter equation with a sophisticated kernel. The 
results indicate that none of the approximations is able to describe the
respone function during or beyond the bipolaronic crossover for the parameters
investigated. In overall, we provide an extensive discussion on when the approximations are valid, and how they fail to describe
the studied exact properties of the chosen model system.
\end{abstract}

\pacs{31.15.xm,31.15.xp,71.10.Fd,71.38-k,71.39.Mx,71.15.Qe}

\date{\today}

\maketitle


\section{Introduction}
\label{Sec:Introduction}
Many-body perturbation theory is one of the most common methodologies used
to study quantum transport problems in which interactions among charge carriers
or between them and other constituents play a significant role. The method is
based on diagrammatic perturbation theory for the non-equilibrium Green's
functions together with a set of standard approximations to describe
the many-body effects~\cite{stefanucci-book}. Although these approximations have
been widely used, and thus their properties explored, in the case of steady-state
transport~\cite{galperin-2007,thygesen-2008,wang-2008,spataru-2009,schmitt-2010},
much less is known on their performance in the explicitly time-dependent
case~\cite{myohanen-2008,myohanen-2009,uimonen-2010,uimonen-2011,myohanen-2012,
khosravi-2012}. This is particularly true for systems with moderate to strong
electron-phonon interactions in which interesting phenomena like bistability and 
hysteresis have been observed~\cite{galperin-2005}. As such phenomena are
typically driven by many-body interactions, it is natural to ask whether or not
the approximate method can describe the relevant physics even qualitatively.
This is the case, for example, with the aforementioned bistability whose
existence has been subject to doubt on the quality of the method
itself~\cite{mitra-2005,alexandrov-2007,galperin-2008,alexandrov-2009, 
dzhioev-2011,albrecht-2012}. The recent efforts to realize more sophisticated,
but computationally more demanding, approximations have enabled addressing these 
questions also in the framework of time-dependent many-body perturbation
theory~\cite{wilner-2013,wilner-2014a,wilner-2014b}. It is important
to study these approximations on a wide scope to understand when they are
predictive and the results can be trusted. Time-dependent many-body perturbation 
theory has also been recently applied to study vibrational effects in ab-initio
charge carrier dynamics in semiconductors e.g.~relaxation processes after a
laser excitation~\cite{marini-2013,sangalli-2015a,sangalli-2015b}.
This has become possible as further simplifications, in particular the
generalized Kadanoff-Baym Ansatz~\cite{lipavsky-1986} (GKBA), have been
developed to keep the approach computationally tractable along with the growing
system sizes. One could also in this manner study time-dependent phenomena in 
realistic molecular systems continuing along the lines of the early studies of 
vibrational effects in photoelectron spectra of molecules~\cite{cederbaum-1974}
In this context, in order to understand the reasons behind the successes or
failures of the methods, it is neccessary to understand the many-body
approximations underlying the additional simplifications such as GKBA.
There are also topical fields in optoelectronics, such as cavity quantum
electrodynamics, and optomechanics in which one deals with formally similar
systems as in the quantum transport case. Time-dependent many-body perturbation
theory has been used in these fields e.g.~to derive time-dependent
density functionals with preliminary results giving an indication of their
quality~\cite{ruggenthaler-2014,pellegrini-2014}. There is however even less
known about the properties of the approximations than in the more established
quantum transport setup.

In this work, we present an extension of a previously introduced numerical
method~\cite{dahlen-2007,stan-2009b} to study time-dependent, inhomogeneous
systems of interacting electrons and phonons. This is also an extension of
the equilibrium formalism which we introduced in our earlier work in
Ref.~\onlinecite{saekkinen-2014a}. Our approach is a variant of time-dependent
many-body perturbation theory based on the Kadanoff-Baym equations
(KBE)~\cite{kadanoff-book}. Here we introduce the relevant equations,
time-dependent many-body approximations, and discuss some of their
characteristic features e.g.~the mean-field, Hartree (H) approximation is shown
to lead to the semi-classical Ehrenfest equations. The time-dependent
partially~\cite{gartner-2006} (Gd) and fully~\cite{wilner-2014a}  (GD)
self-consistent Born approximations are introduced to study correlation effects
beyond the mean-field level. These many-body approximations are in particular
suited to study time-dependent quantum transport with electron-phonon
interactions as they are particle number conserving in the sense of
Baym~\cite{baym-1961,baym-1962}. In the present work, the method is instead
applied to a finite system since this allows us to assess its performance by
comparing the approximate results to an exact solution. Although the method can
handle complex time-dependent perturbations, we restrict ourselves here to
linear response functions obtained by time-propagating the Kadanoff-Baym
equations. The density response function $\delta n/\delta v$ obtained in this
manner is equivalent to a solution of the Bethe-Salpeter equation (BSE) with an 
integral kernel which is a functional derivative $\delta \Sigma/\delta G$ of the
self-energy $\Sigma$ with respect to the electron propagator
$G$~\cite{kwong-2000,dahlen-2007,saekkinen-2012,balzer-2012}.The kernel
therefore consists of dressed propagators and is fully frequency-dependent in
the Born approximations. This level of approximation has, to our best knowledge, 
not yet been reached in the standard frequency-domain approach even for the much 
studied purely electronic systems~\cite{onida-2002,romaniello-2009,
sangalli-2011,zhang-2013}. Moreover, as the phonon  propagator is determined by 
the electronic response function, by comparing to the equilibrium phonon 
propagator we are able to comment on whether or not this additional level of 
sophistication amounts to improved results. Lastly, we would like to note that 
although this paper is geared towards electrons and phonons, the method is in 
fact applicable to a variety of systems of interacting fermions and bosons 
e.g.~electron-photon models (Rabi~\cite{rabi-1936,rabi-1937}) and
electron-plasmon models (Lundqvist~\cite{lundqvist-1967}).

As the application, we study a homogeneous, two-site, two-electron Holstein
model which is a standard model describing interacting electrons and
phonons~\cite{holstein-2000}. This continues along the lines of our prior work
in Ref.~\onlinecite{saekkinen-2014a} in which we focused on ground-state
properties and studying the localizing effect of the electron-phonon interaction
by comparing the many-body approximations against exact benchmark results. As
a result, we found that the self-energy approximations gave rise to spontaneous
symmetry-breaking characterizable by an asymmetric electron density and nuclear 
displacement. The symmetry-broken solutions as well as solutions obtained by
enforcing symmetry were analyzed with the help of total energies, energy
components, and natural occupation numbers. It was concluded that the
symmetry-breaking can be seen physically to mimic the bipolaronic crossover of
the underlying system in which two nearly free electrons form a bound pair with
an accompanying nuclear displacement. Moreover, out of the symmetric solutions,
only the fully self-consistent Born approximation showed evidence of partially
describing this crossover. Here we instead investigate the equilibrium electron
and phonon propagators, and linear response functions of the same system using
time-dependent many-body perturbation theory. The equilibrium propagators are
studied in frequency-domain which gives a more detailed view to the properties
of the approximations, and allows us to re-evaluate the physical picture
obtained from the various energies. The linear response calculations on the
other hand allow us to understand better the nature of the symmetric and
asymmetric solutions found in our earlier work. In particular, we show that
they are equivalent to the equilibrium solutions of the semi-classical equations
of the Dicke model~\cite{dicke-1954} in which the appearance of the asymmetric
solution represents the super-radiant phase transition in the thermodynamic
limit~\cite{hepp-1973,wang-1973,emery-2003}. This transition moreover appears as
a bifurcation leading to instability of the symmetric solution which in a finite 
system is not in agreement with the exact solution. One of the open questions addressed in this work concerns the stability of the symmetric and asymmetric 
solutions when going beyond the mean-field approximation. In particular, we
answer to the question whether or not the symmetric solution retains its
stability in the Born approximations. Once stable solutions have been
identified, we turn our attention to the linear response functions which are
used to assess how the approximations describe the system reacting
to a weak perturbation. There is a lot of systematic work on static
i.e.~zero-frequency susceptibilities of either finite clusters~\cite{levine-1991},
or extended finite~\cite{marsiglio-1990} and infinite~\cite{freericks-1993,freericks-1994} dimensional systems with a focus on
e.g~charge-density wave phase transition temperatures. Here we thus extend these 
studies beyond the static case by considering fully frequency-dependent response functions of a finite system. 

The paper is organized as follows. In Sec. \ref{Sec:Theory}, we introduce
our method: time-dependent many-body perturbation for electrons
interacting with phonons. The method is applied in this work to the model system
introduced in Sec.~\ref{Sec:Model}. The results containing both the
equilibrium electron addition and removal, as well as neutral excitation
spectra are presented, analyzed and discussed in Sec.~\ref{Sec:Results}.
The conclusions and an outlook are given in Sec.~\ref{Sec:Conclusions}, and some
more technical details are presented in App.~\ref{App:HartreeTdscfEhrenfest}
and~\ref{App:HartreeDensityResponse}.


\section{Theory}
\label{Sec:Theory}


\subsection{Hamiltonian}
\label{Sec:Theory:Hamiltonian}
In the present work, we introduce the central concepts of time-dependent
many-body perturbation theory for systems of electrons interacting with phonons.
Although we do not discuss here electron-electron interactions, they could be
included without additional conceptual difficulties. The time-dependent
Hamiltonian operator is then given by
\begin{align*}
   \hat{H}(z)
   =&\sum_{p}\omega_{p}(z)\hat{a}_{p}^{\dagger}\hat{a}_{p}
   +\sum_{p}\big(f_{p}(z)\hat{a}_{p}^{\dagger}+f_{p}^{*}(z)\hat{a}_{p}\big)
   \notag\\
   &+\sum_{ij}h_{ij}(z)\hat{c}^{\dagger}_{i}\hat{c}_{j} \notag\\
   &+\sum_{ij}\sum_{p}\big(
   m_{jk}^{p}(z)\hat{a}^{\dagger}_{p}
   +m_{kj}^{p{*}}(z)\hat{a}_{p}\big)
   \hat{c}^{\dagger}_{j}\hat{c}_{k}
   \, ,
\end{align*}
where the properties of the system are encoded in the phonon frequencies
$\omega_{p}$, generalized forces $f_{p}$, elements of the electron one-body
Hamiltonian $h_{ij}$, and electron-phonon interaction elements $m_{ij}^{p}$.

\begin{figure}
   \centering
   \includegraphics[width=0.3\textwidth]{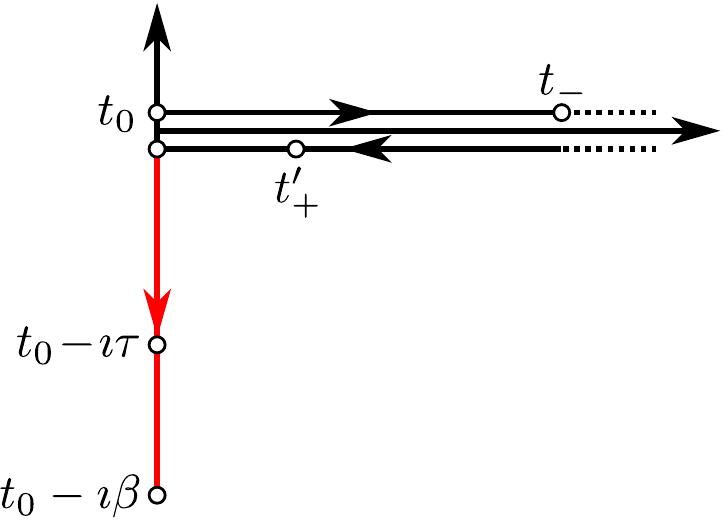}
   \caption{\label{Fig:Contour}
   The extended Keldysh contour which consists of the vertical, imaginary-time
   track responsible for the initial equilibrium preparation, and of
   the horizontal forward ($-$) and backward ($+$) real-time tracks related to
   the real-time time-evolution. (color online)}
\end{figure}

These quantities all depend on a time-argument lying on the extended Keldysh
contour~\cite{stefanucci-book} shown in Fig.~\ref{Fig:Contour}. The Hamiltonian
operator is however the same on the forward ($-$) and backward ($+$) branches,
and independent of the contour time on the vertical equilibrium ($M$) track
as in our prior work in Ref.~\onlinecite{saekkinen-2014a}. An explicit
time-dependence allows one to realize a variety of physical scenarios from
electrons and nuclei driven by electromagnetic fields to more abstract
simulations based on an interaction quench. In the present
work, we however focus on another type of time-dependence arising from the
choice of the initial state.

The electrons and phonons are described in second quantization with annihilation
$\hat{c}_{i}$, $\hat{a}_{p}$ and creation $\hat{c}^{\dagger}_{i}$,
$\hat{a}_{p}^{\dagger}$ operators which obey canonical anti-commutation and
-commutation relations, respectively. In order to facilitate a compact
presentation of the many-body perturbation theory, we further introduce
the self-adjoint phonon operators
\begin{align*}
   \hat{\phi}_{1,p}
   &\equiv\big(\hat{a}^{\dagger}_{p}+\hat{a}_{p}\big)/\sqrt{2}\, ,
   &\hat{\phi}_{2,p}
   &\equiv\imath\big(\hat{a}^{\dagger}_{p}-\hat{a}_{p}\big)/\sqrt{2}\, ,
\end{align*}
to which we associate a collective index $P\equiv\{\varsigma_{p}\in\{1,2\},p\}$
so that we can write their commutation relation compactly as
\begin{align*}
   [\hat{\phi}_{P},\hat{\phi}_{Q}]
   =\alpha_{PQ}
   \, ,
\end{align*}
where $\alpha_{1p,1q}=\alpha_{2p,2q}=0$ and
$\alpha_{1p,2q}=-\alpha_{2q,1p}=\imath\delta_{pq}$. These operators can be
physically understood as components of the displacement ($\hat{\phi}_{1p}$) and
momentum ($\hat{\phi}_{2p}$) operators. They allow us to rewrite the Hamiltonian
operator as
\begin{align}
\label{Eq:GeneralHamiltonian}
   \hat{H}(z)
   =&\sum_{PQ}\Omega_{PQ}(z)\op{\phi}_{P}\op{\phi}_{Q}
   +\sum_{P}F_{P}(z)\op{\phi}_{P}
   \notag\\
   &+\sum_{ij}h_{ij}(z)\op{c}^{\dagger}_{i}\op{c}_{j}
   \notag\\
   &+\sum_{ij}\sum_{P}M_{ij}^{P}(z)\op{\phi}_{P}\op{c}^{\dagger}_{i}\op{c}_{j}
   \, ,
\end{align}
where the phonon frequencies, generalized forces, and electron-phonon
interaction are incorporated into
\begin{align*}
   F_{p\varsigma_{p}}(z)
   &\equiv \delta_{\varsigma_{p},1}\big(f_{p}(z)+f^{*}_{p}(z)\big)/\sqrt{2}
   \notag\\
   &-\imath\delta_{\varsigma_{p},2}\big(f_{p}(z)-f^{*}_{p}(z)\big)/\sqrt{2}
   \, ,\\
   \Omega_{p\varsigma_{p},q\varsigma_{q}}(z)
   &\equiv\omega_{p}(z)(\delta_{pq}\delta_{\varsigma_{p}\varsigma_{q}}
   +\alpha_{p\varsigma_{p},q\varsigma_{q}})/2
   \, ,\\
   M_{jk}^{p\varsigma_{p}}(z)
   &\equiv\delta_{\varsigma_{p},1}\big(m_{jk}^{p}(z)
   +m_{kj}^{p{*}}(z)\big)/\sqrt{2}
   \notag\\
   &-\imath\delta_{\varsigma_{p},2}\big(m_{jk}^{p}(z)
   -m_{kj}^{p{*}}(z)\big)/\sqrt{2}
   \, ,
\end{align*}
which are to be understood in this work to represent elements of a vector,
matrix, and a vector of matrices, respectively. The one-body electron
Hamiltonian elements are also to be understood as elements of a matrix.
In the following an overhead arrow denotes a vector ($\V{F}$), boldfaced symbols
matrices ($\M{\Omega},\M{h}$), and a combination of these two a vector of
matrices ($\T{M}$), while $\tr$ denotes a matrix trace.

\subsection{Many-Body Perturbation Theory}
\label{Sec:Theory:ManyBodyPerturbationTheory}
The central quantities of many-body perturbation theory of interacting
electrons and phonons are the phonon field expectation value, and the phonon and
electron propagators defined as
\begin{align}
   \phi_{P}(z)
   &\equiv\frac{1}{\Z}
   \Tr\bigg[\toop\Big\{e^{-\imath\int_{C}\!d\bar{z}\;\hat{H}(\bar{z})}
   \hat{\phi}_{P}(z)\Big\}\bigg]
   \, ,\notag\\
   D_{PQ}(z;z')
   &\equiv \frac{1}{\imath\Z}
   \Tr\bigg[\toop\Big\{e^{-\imath\int_{C}\!d\bar{z}\;H(\bar{z})}
   \Delta\hat{\phi}_{P}(z)\Delta\hat{\phi}_{Q}(z')\Big\}\bigg]
\label{Eq:DefD}
   \, ,\\
   G_{ij}(z;z')
   &\equiv \frac{1}{\imath\Z}
   \Tr\bigg[\toop\Big\{e^{-\imath\int_{C}\!d\bar{z}\;H(\bar{z})}
   \hat{c}_{i}(z)\hat{c}^{\dagger}_{j}(z')\Big\}\bigg]
\label{Eq:DefG}
   \, ,
\end{align}
where $\Delta\hat{\phi}_{P}\equiv\hat{\phi}_{P}-\phi_{P}$ is a fluctuation
operator, $\Z\equiv\Tr[e^{-\imath\int\!dz\;\hat{H}(z)}]$
the partition function, $\Tr$ the trace over a complete
set of quantum states, and $\toop$ is the time-ordering operator on a Keldysh
time-contour $C$ of Fig.~\ref{Fig:Contour} acting on operators given in
the Schr\"{o}dinger picture but having time-arguments $z,z'$ for book-keeping
reasons~\cite{stefanucci-book}. These objects have a closed form perturbation
expansion obtained using  Wick's theorem and re-summing all terms into two
electron and phonon propagator line irreducible contributions. This leads to
the equations
\begin{subequations}
\label{Eq:DysonEquations}
\begin{align}
   \V{\phi}(z)
   &=\int_{C}\!d\bar{z}\; \M{d}(z;\bar{z})\big(\V{F}(\bar{z})
   -\imath\tr\big(\T{M}(\bar{z})\M{G}(\bar{z};\bar{z}^{+})\big)
   \, ,\\
\label{Eq:DysonEquationPhononPropagator}
   \M{D}(z;z')
   &=\M{d}(z;z')+\int_{C}\!d\bar{z}d\bar{z}'\;
   \M{d}(z;\bar{z})
   \M{\Pi}(\bar{z};\bar{z}')
   \M{D}(\bar{z}';z')
   \, ,\\
   \M{G}(z;z')
   &=\M{g}(z;z')+\int_{C}\!d\bar{z}d\bar{z}'\; \M{g}(z;\bar{z})
   \M{\Sigma}(\bar{z};\bar{z}')
   \M{G}(\bar{z}';z')
   \, ,
\end{align}
\end{subequations}
where $g$ and $d$ denote the non-interacting electron and phonon propagators
defined by Eqs.~\eqref{Eq:DefG}~and~\eqref{Eq:DefD} in the absence of the
electron-phonon interaction. The integral kernels $\Sigma\equiv\Sigma[G,D]$ and
$\Pi\equiv\Pi[G,D]$ are non-local one-body potentials known as electron and
phonon self-energies. These self-energies contain information on interactions of
the system, as well as the external driving induced by the generalized force $F$.
The non-interacting electron and phonon propagators are given respectively by
\begin{align*}
   \M{g}(z;z')
   &=-\imath\M{U}(z,t_{0})\Big(
   \M{\theta}(z,z')-\M{f}_{+}\big(\beta\M{h}^{M}\big)
   \Big)\M{U}(t_{0},z')
   \, ,\\
   \M{d}(z;z')
   &=-\imath\M{\alpha}\M{V}(z,t_{0})\Big(
   \M{\theta}(z,z')+\M{f}_{-}\big(\beta\M{\tilde{\Omega}}^{M}\M{\alpha}\big)
   \Big)\M{V}(t_{0},z')
   \, ,
\end{align*}
where $\M{\theta}\equiv\theta\M{1}$ with $\theta$ being the Heaviside function
and $\M{1}$ the identity matrix, $\beta$ is the inverse temperature,
$f_{\pm}$ denote the Fermi-Dirac ($+$) and Bose-Einstein ($-$) distribution
functions, $\M{\tilde{\Omega}}^{M}\equiv\M{\tilde{\Omega}}(t_{0}-\imath\tau)$
independent of $\tau$ is the Matsubara component of
\begin{align*}
   \M{\tilde{\Omega}}(z)
   \equiv\M{\Omega}(z)+\M{\Omega}^{T}(z)
   \, .
\end{align*}
Finally, we introduced the time-evolution matrices as solutions to
\begin{align*}
   \imath\partial_{z}\M{U}(z,z')
   &=\M{h}(z)\M{U}(z,z')
   \, ,\\
   -\imath\partial_{z'}\M{U}(z,z')
   &=\M{U}(z,z')\M{h}(z')
   \, ,\\
   \imath\partial_{z}\M{V}(z,z')
   &=\M{\tilde{\Omega}}(z)\M{\alpha}\M{V}(z,z')
   \, ,\\
   -\imath\partial_{z'}\M{V}(z,z')
   &=\M{V}(z,z')\M{\tilde{\Omega}}(z)\M{\alpha}
   \, ,
\end{align*}
with the initial conditions $\M{U}(t_{0},t_{0})=\M{V}(t_{0},t_{0})=\M{1}$.

In our earlier work in Ref.~\onlinecite{saekkinen-2014a}, we introduced our
implementation of the equilibrium Matsubara formalism obtained by choosing
time-arguments $z=t_{0}-\imath\tau,z'=t_{0}-\imath\tau'$ on the imaginary track.
Here we focus on an extension of this formalism to time-dependent cases in which
it is more natural to differentiate Eqs.~\eqref{Eq:DysonEquations} with respect
to the first contour time in order to arrive at the equations of motion
\begin{align}
\label{Eq:EquationsOfMotion}
   &\big(\imath\M{\alpha}\partial_{z}
   -\M{\tilde{\Omega}}(z)\big)\V{\phi}(z)
   \notag\\
   &=\V{F}(z)
   -\imath\tr\big(\T{M}(z)\M{G}(z;z^{+})\big)
   \, , \\
   &\big(\imath\M{\alpha}\partial_{z}
   -\M{\tilde{\Omega}}(z)\big)\M{D}(z;z')
   \notag\\
   &=\M{\delta}(z,z')
   +\int_{C}\!dz\; \M{\Pi}(z;\bar{z})\M{D}(\bar{z};z')
   \, ,\\
   &\big(\M{\imath}\partial_{z}-\M{h}(z)\big)\M{G}(z;z')
   \notag\\
   &=\M{\delta}(z,z')
   +\int_{C}\!dz\; \M{\Sigma}(z;\bar{z})\M{G}(\bar{z};z')
   \, ,
\end{align}
where $\M{\imath}=\imath\M{1}$. These equations together with their conjugate
equations obtained by differentiating with respect to the second time-argument
of the propagators form a closed set of the equations which can be solved once
an approximation for the many-body part of the self-energy has been fixed.


\subsection{Self-Energies}
\label{Sec:Theory:SelfEnergyApproximations}
\begin{figure}
   \centering
   \includegraphics[scale=0.7]{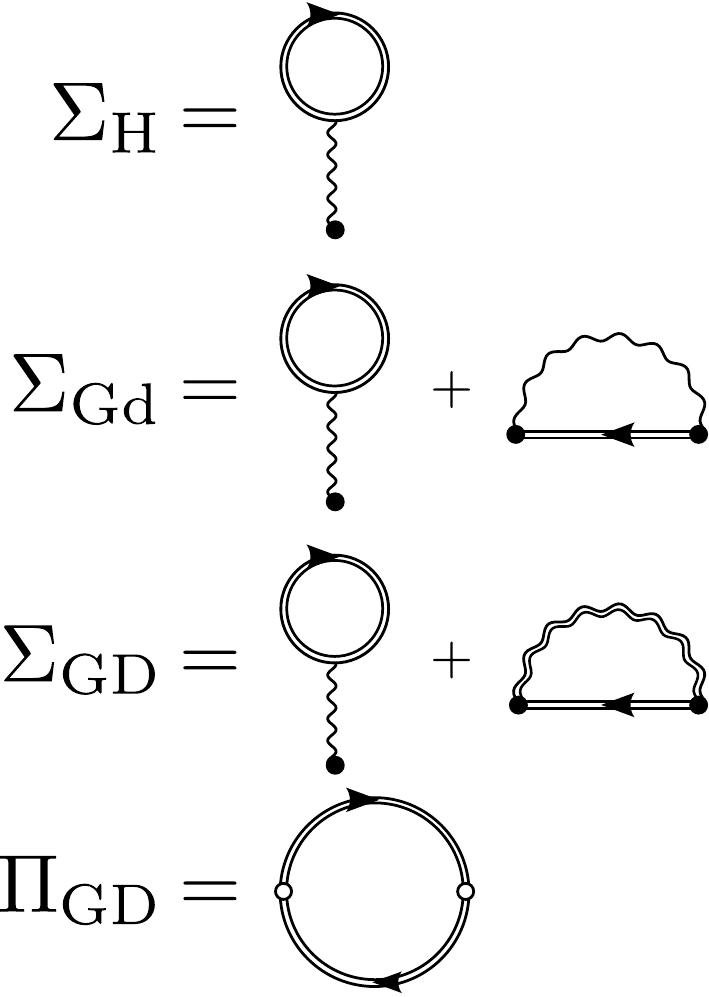}
   \caption{\label{Fig:Selfenergy}The Hartree (H), partially self-consistent
   (Gd), and fully self-consistent (GD) Born self-energies summarize
   the many-body approximations used in this work. A two-fold line with an arrow
   indicates a dressed electron propagator, while single and two-fold wiggly
   lines represent bare and dressed phonon propagators, respectively.
   An open circle and a closed circle represent a connection for a phonon and
   an electron propagator, respectively.}
\end{figure}

The self-energy $\Sigma$, as noted above, contains both a contribution arising
from the generalized force $F_{P}(z)$, as well as a part induced by
the electron-phonon interactions. The phonon propagator, being defined
in terms of fluctuation operators, is not directly influenced by this force,
instead it appears in the electron self-energy and can be handled by writing
the self-energy as
\begin{align*}
   \Sigma_{ij}(z;z')
   &=\delta(z,z')v_{n,ij}(z)+\Sigma_{\mathrm{MB},ij}(z;z')
\end{align*}
where we introduced the potential
\begin{align*}
   v_{n,ij}(z)
   &\equiv\sum_{PQ}M_{ij}^{P}(z)\int_{C}\!d\bar{z}\;
   d_{PQ}(z,\bar{z})F_{Q}(\bar{z})
\end{align*}
which represents the classical potential induced by nuclei
experiencing a generalized force $F_{Q}$. The many-body self-energy, denoted by
$\mathrm{MB}$, is then subject to approximation. The approximations used here,
and introduced earlier in Ref.~\onlinecite{saekkinen-2014a}, are summarized
diagrammatically in Fig.~\ref{Fig:Selfenergy}. The approximate electron
self-energies consists of the Hartree (H) and Fock (F) diagrams. The Hartree
diagram can be written as
\begin{align*}
   \Sigma_{\mathrm{H}}&[G]_{ij}(z;z')
   =\delta(z,z')v_{\mathrm{H}}[G]_{ij}(z)
   \, ,
\end{align*}
where the time-local Hartree potential is given by
\begin{align}
\label{Eq:HartreePotentialA}
   v_{\mathrm{H}}[G]_{ij}(z)
   &=-\imath\sum_{\substack{kl\\PQ}}M_{ij}^{P}(z)
   \notag\\
   &\times\int_{C}\!d\bar{z}\;
   d_{PQ}(z,\bar{z})M_{kl}^{Q}(\bar{z})G_{lk}(\bar{z};\bar{z}^{+})
   \, .
\end{align}
or alternatively by
\begin{align}
\label{Eq:HartreePotentialB}
   v_{\mathrm{H}}[G]_{ij}(z)
   &=\sum_{P}M_{ij}^{P}(z)\phi_{P}(z)
   -v_{n,ij}(z)
   \, ,
\end{align}
which follows from the equation of motion for the non-interacting phonon
propagator. Electron self-energy terms beyond Hartree contribute to the
exchange-correlation, many-body self-energy
\begin{align*}
   \Sigma_{\mathrm{xc},ij}(z;z')
   &\equiv\Sigma_{\mathrm{MB},ij}(z;z')-\Sigma_{\mathrm{H},ij}(z;z')
   \, ,
\end{align*}
whose lowest order diagram is the Fock diagram
\begin{align*}
   \Sigma_{\mathrm{F}}&[G,D]_{ij}(z;z')
   \notag\\
   &= \imath\sum_{kl,PQ}
   M_{ik}^{P}(z)M_{lj}^{Q}(z')
   D_{PQ}(z;z') G_{kl}(z;z')
   \, , 
\end{align*}
which is a time-nonlocal memory term describing single-phonon
absorption/emission processes. The only phonon self-energy diagram used
in this work is the bubble diagram 
\begin{align*}
   \Pi_{\mathrm{B}}&[G]_{PQ}(z;z')
   \notag\\
   &=-\imath\sum_{ij,kl}
   M_{ij}^{P}(z)M_{kl}^{Q}(z')G_{li}(z';z)G_{jk}(z;z') \, ,
\end{align*}
which describes simple phonon induced electron-hole excitation processes.\\

\begin{figure}
   \centering
   \includegraphics[scale=0.7]{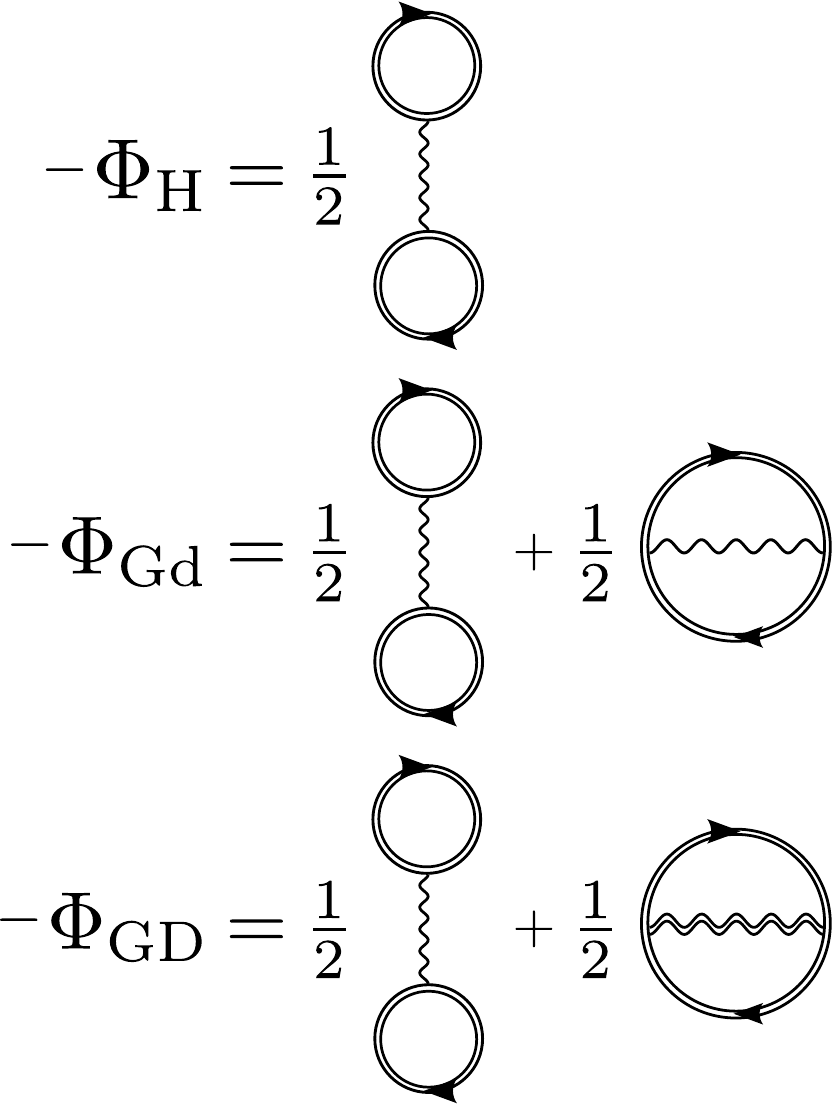}
   \caption{\label{Fig:PhiFunctional}
   The $\Phi$-functionals for the Hartree (H), partially self-consistent (Gd),
   and fully self-consistent (GD) Born approximations. A two-fold line with an 
   arrow indicates a dressed electron propagator, while single and two-fold
   wiggly lines represent bare and dressed phonon propagators, respectively.
   Note that the minus sign on the left hand side arises due to the loop
   rule~\cite{stefanucci-book}.}
\end{figure}

The many-body self-energies and their abbreviations used throughout the text
are summarized in the list below.
\begin{description}
\item[H]
The Hartree approximation consists of approximating the electron self-energy
with the Hartree diagram
\begin{align*}
   \Sigma_{\mathrm{H}}&[G]_{ij}(z;z')
   =\delta(z,z')v_{\mathrm{H}}[G]_{ij}(z)
   \, ,
\end{align*}
and neglecting the phonon self-energy. This is
a mean-field approximation in which electrons feel only the classical potential
due to nuclei. The resulting Hartree equations
\begin{subequations}
\label{Eq:HartreeEquations}
\begin{align}
   \imath\frac{d}{dz}\M{G}(z;z^{+})
   &=\big[\M{h}(z)+\M{v}_{n}(z)
   \notag\\
   &+\M{v}_{\mathrm{H}}(z),\M{G}(z;z^{+})\big]
   \, ,\\
   \imath\M{\alpha}\partial_{z}\V{\phi}(z)
   &=\M{\tilde{\Omega}}(z)\V{\phi}(z)+\V{F}(z)
   \notag\\
   &-\imath\tr\big(\T{M}\M{G}(z;z^{+})\big)
   \, ,
\end{align}
\end{subequations}
can be shown to be equivalent to the semi-classical Ehrenfest equations,
see App.~\ref{App:HartreeTdscfEhrenfest}.

\item[Gd]
The partially self-consistent Born approximation amounts to approximating
the electron many-body self-energy with
\begin{align*}
   \Sigma_{\mathrm{Gd}}[G]_{ij}(z;z')
   &\equiv\Sigma_{\mathrm{H}}[G]_{ij}(z;z')
   \notag\\
   &+\Sigma_{\mathrm{F}}[G,d]_{ij}(z;z') \, ,
\end{align*}
where $d$ is the bare phonon propagator obtained by putting the phonon
self-energy to zero. This amounts to saying that that the nuclei are unaffected
by the electronic particle-hole excitations.

\item[GD]
The fully self-consistent Born approximation is defined by writing
the electron many-body self-energy as
\begin{align*}
   \Sigma_{\mathrm{GD}}[G,D]_{ij}(z;z')
   &\equiv\Sigma_{\mathrm{H}}[G]_{ij}(z;z')
   \notag\\
   &+\Sigma_{\mathrm{F}}[G,D]_{ij}(z;z') \, 
\end{align*}
while the phonon self-energy is given by 
\begin{align*}
   \Pi_{\mathrm{GD}}[G]_{PQ}(z;z')
   &\equiv\Pi_{\mathrm{B}}[G]_{PQ}(z;z')
   \, .
\end{align*}
Note that although we dress the phonon propagator in the Fock diagram,
one should not use a dressed propagator in the Hartree diagram as it leads
to double-counting of terms in the perturbation
expansion~\cite{hedin-1965,stefanucci-book}.
\end{description}
These approximations are all $\Phi$-derivable, that is the corresponding
self-energies can be obtained as the functional derivatives
\begin{align*}
   \Sigma_{ij}[G,D](z;z')
   &=\frac{\delta\Phi[G,D]}{\delta G_{ji}(z';z)}
   \, ,\\
   \Pi_{PQ}[G,D](z;z')
   &=-2\frac{\delta\Phi[G,D]}{\delta D_{QP}(z';z)}\bigg|_{S}
   \, ,
\end{align*}
of an approximate $\Phi$-functional which are shown in
Fig.~\ref{Fig:PhiFunctional}. Note that the subscript $S$ refers to
a symmetrized functional derivative, see\cite{stefanucci-book}.
The $\Phi$-derivability of the electron self-energy together with
self-consistency in the electron propagator guarantee gauge invariance and
consequently fulfillment of the electron density conservation
law~\cite{baym-1961,baym-1962}.

\subsection{Kadanoff-Baym Equations}
\label{Sec:Theory:KadanoffBaymEquations}
The equations of motion of Eqs.~\eqref{Eq:EquationsOfMotion} are customary
solved by projecting the propagators to different parts of the Keldysh contour
by choosing the time-arguments appropriately~\cite{stefanucci-book}.
This procedure leads to the greater ($>$), lesser ($<$), left ($\lceil$),
right ($\rceil$), and Matsubara ($M$) components
\begin{align*}
   a^{\gtrless}(t;t')
   &\equiv a(t_{\pm};t_{\mp}')
   \, ,\\
   a^{\rceil}(t;\tau)
   &\equiv a(t;t_{0}-\imath\tau)
   \, ,\\
   a^{\lceil}(\tau;t)
   &\equiv a(t_{0}-\imath\tau;t)
   \, ,\\
   a^{M}(\tau;\tau')
   &\equiv a(t_{0}-\imath\tau;t_{0}-\imath\tau')
   \, ,
\end{align*}
where the subscript $\mp$ denotes a time evaluated on the forward/backward
branches of the contour, and $a$ is a function in the space of Keldysh
functions~\cite{stefanucci-book}. The Keldysh components of the phonon and
electron propagator obey the symmetries
\begin{align*}
   G^{\gtrless}_{ij}(t;t')
   &=-\big[G_{ji}^{\gtrless}(t';t)\big]^{*}
   \, ,\\
   G_{ij}^{\lceil}(\tau;t)
   &=\big[G_{ji}^{\rceil}(t;\beta-\tau)\big]^{*}
   \, ,\\
   D_{PQ}^{\gtrless}(t;t')
   &=-\big[D_{QP}^{\gtrless}(t';t)\big]^{*}
   =-\big[D_{PQ}^{\lessgtr}(t;t')\big]^{*}
   \, ,\\
   D_{PQ}^{\lceil}(\tau;t)
   &=\big[D_{QP}^{\rceil}(t;\beta-\tau)\big]^{*}
   =\big[D_{PQ}^{\lceil}(\beta-\tau;t)\big]^{*}
   \, ,
\end{align*}
where the additional symmetries of the phonon propagator are due to
the symmetry $D_{PQ}(z;z')=D_{QP}(z',z)$. The equations of motion obtained
by taking all possible components of the contour-time equations of motion
form a set of non-linear integro-differential equations of motion known as
the Kadanoff-Baym equations~\cite{kadanoff-book}. The symmetries of
the propagator however imply that we only need equations of motion for
the greater, lesser, and right components where the first two are required
for times $t\geq t'$. The relevant equations of motion are then
\begin{align*}
   \imath\partial_{t}\M{G}^{\gtrless}(t;t')
   &=\M{h}_{\mathrm{eff}}(t)\M{G}^{\gtrless}(t;t')
   +\M{I}^{\gtrless}[\Sigma_{\mathrm{xc}},G](t;t')
   \, ,\\
   \imath\partial_{t}\M{G}^{\rceil}(t;\tau)
   &=\M{h}_{\mathrm{eff}}(t)\M{G}^{\rceil}(t;\tau)
   +\M{I}^{\rceil}[\Sigma_{\mathrm{xc}},G](t;t')
   \, ,\\
   \imath\partial_{t}\M{D}^{\gtrless}(t;t')
   &=\M{\alpha}\Big(\M{\tilde{\Omega}}(t)\M{D}^{\gtrless}(t;t')
   +\M{I}^{\gtrless}[\Pi,D](t;t')\Big)
   \, ,\\
   \imath\partial_{t}\M{D}^{\rceil}(t;\tau)
   &=\M{\alpha}\Big(\M{\tilde{\Omega}}(t)\M{D}^{\rceil}(t;\tau)
   +\M{I}^{\rceil}[\Pi,D](t;\tau)\Big)
   \, ,
\end{align*}
for off time-diagonal and 
\begin{align*}
   \imath\frac{d}{dt}\M{G}^{\gtrless}(t;t)
   &=\M{h}_{\mathrm{eff}}(t)\M{G}^{\gtrless}(t;t)
   +\M{I}^{\gtrless}[\Sigma_{\mathrm{xc}},G](t;t)
   +\mathrm{h.c.}
   \, ,\\
   \imath\frac{d}{dt}\M{D}^{\gtrless}(t;t)
   &=\M{\alpha}\Big(\M{\tilde{\Omega}}(t)\M{D}^{\gtrless}(t;t)
   +\M{I}^{\gtrless}[\Pi,D](t;t)\Big)+\mathrm{h.c.}
   \, ,
\end{align*}
where $\mathrm{h.c.}$ denotes the Hermitian conjugate, for on time-diagonal
time-propagation. Here we introduced the effective one-body electron
Hamiltonian
\begin{align*}
   \M{h}_{\mathrm{eff}}(t)
   &\equiv\M{h}(t)+\M{v}_{n}(t)+\M{v}_{\mathrm{H}}(t)
   \, ,
\end{align*}
as well as the collision integrals
\begin{align*}
   I^{\gtrless}[a,b](t;t')
   &\equiv \big[a^{\rceil}\star b^{\lceil}\big](t;t')
   \notag\\
   &+\big[a^{R}\bullet b^{\gtrless}\big](t;t')
   +\big[a^{\gtrless}\bullet b^{A}\big](t;t')
   \, ,\\
   I^{\rceil}[a,b](t;\tau)
   &\equiv\big[a^{\rceil}\star b^{M}\big](t;\tau)
   +\big[a^{R}\bullet b^{\rceil}\big](t;\tau)
   \, .
\end{align*}
with the bullets and stars denoting convolution integrals of the form
\begin{align*}
   \big[a\bullet b](t;t')
   &=\int_{t_{0}}^{\infty}\!d\bar{t}\; a(t,\bar{t})b(\bar{t},t') \, ,
   \\
   \big[a\star b](t;t')
   &=-\imath\int_{0}^{\beta}\!d\tau\; a(t,\tau)b(\tau,t') \, ,
\end{align*}
where $a$ and $b$ are possibly matrix valued functions on the Keldysh contour.
The Hartree potential appearing in the effective Hamiltonian can be evaluated
using Eq.~\eqref{Eq:HartreePotentialB} instead of
Eq.~\eqref{Eq:HartreePotentialA} by taking advantage of the equation of motion 
\begin{align*}
   \imath\partial_{t}\V{\phi}(t)
   &=\M{\alpha}\Big(\M{\tilde{\Omega}}(t)\V{\phi}(t)
   \notag\\
   &+\V{F}(t)-\imath\tr\big(\T{M}(t) \M{G}^{<}(t;t)\big)\Big)
   \, ,
\end{align*}
for the real-time $\phi_{P}(t)\equiv\phi_{P}(t_{\pm})$ phonon field expectation
value. The Kadanoff-Baym equations, including the equation above, then form a
closed set of equations which, when supplemented with the initial conditions
\begin{align*}
   G_{ij}^{\gtrless}(t_{0};t_{0})
   &=G_{ij}^{M}(0^{\pm})
   \, ,\\
   G_{ij}^{\rceil}(t_{0};\tau)
   &=G_{ij}^{M}(-\tau)
   \, ,\\
   D_{PQ}^{\gtrless}(t_{0};t_{0})
   &=D_{PQ}^{M}(0^{\pm})
   \, ,\\
   D_{PQ}^{\rceil}(t_{0};\tau)
   &=D_{PQ}^{M}(-\tau)
   \, ,\\
   \phi_{P}(t_{0})
   &=\phi_{P}^{M}
   \, ,
\end{align*}
given by the equilibrium Matsubara components introduced in
Ref.~\onlinecite{saekkinen-2014a}, can be solved on a computer by
time propagation~\cite{stan-2009b}.


\section{Model}
\label{Sec:Model}
Our model system is a two-site Holstein model~\cite{
schmidt-1987,feinberg-1990,ranninger-1992,alexandrov-1994,
marsiglio-1995,demello-1997,crljen-1998,rongsheng-2002,qingbao-2005,
paganelli-2008a,paganelli-2008b,firsov-1997,chatterjee-2002}
which can be viewed as a minimal representation of a system in which electrons 
move between two molecules, so that they are coupled to the local vibrational 
modes of these molecules. In the case of two identical molecules, we find that 
only the relative displacement couples to the electron density difference
between the molecules, and thus the Hamiltonian operator for the isolated system 
reduces to
\begin{align*}
   \hat{H}^{M}
   &\equiv\omega_{0}\hat{a}^\dagger\hat{a}
   \notag\\
   &-t_{\mathrm{kin}}\sum_{\sigma}\big(
   \hat{c}_{1\sigma}^{\dagger}\hat{c}_{2\sigma}
   +\hat{c}_{2\sigma}^{\dagger} \hat{c}_{1\sigma}\big)
   \notag\\
   &-\frac{g}{\sqrt{2}}(\hat{a}^{\dagger}+\hat{a})\sum_{\sigma}
   \big(\hat{n}_{1\sigma}-\hat{n}_{2\sigma}\big)
   \, ,
\end{align*}
where $\hat{a}$ and $\hat{a}^{\dagger}$ annihilate and create a phonon to
the relative displacement mode, $\hat{c}_{i\sigma}$ is the electronic operator
that annihilates an electron of spin $\sigma$ at site $i$, and
$\hat{n}_{i\sigma}\equiv\hat{c}^{\dagger}_{i\sigma}\hat{c}_{i\sigma}$ is
the electron density operator at site $i$. The parameters $\omega_{0}$,
$t_{\mathrm{kin}}$ and $g$ characterize the bare vibrational frequency,
inter-site hopping and local electron-phonon interaction strength, respectively.
This Hamiltonian corresponds to the relative Hamiltonian of
Ref.~\onlinecite{saekkinen-2014a} which is chosen here over the full Hamiltonian 
due to both its simplicity and computational reasons. The system can be probed 
with external time-dependent fields which are described with the Hamiltonian operator 
\begin{align*}
   \hat{H}(t)
   &\equiv\hat{H}^{M}
   +f(t)\big(\hat{a}^{\dagger}+\hat{a}\big)
   +\sum_{i\sigma}v_{i}(t)\hat{n}_{i\sigma}
   \, ,
\end{align*}
where $f$ and $v_{i}$ describe amplitudes of the external fields acting on
the nuclei and electrons, respectively. The displacement and momentum operators,
defined in this model as $\hat{u}\equiv(\hat{a}^{\dagger}+\hat{a})/\sqrt{2}$ and
$\hat{p}\equiv\imath(\hat{a}^{\dagger}-\hat{a})/\sqrt{2}$, allow us to rewrite
the Hamiltonian operator as
\begin{subequations}
\label{Eq:HamiltonianModelSystem}
\begin{align}
   \hat{H}(t)
   =&\hat{H}^{M}
   +\sqrt{2}f(t)\hat{u}
   +\sum_{i\sigma}v_{i}(t)\hat{n}_{i\sigma}
   \, ,\\
   \hat{H}^{M}
   =&\frac{\omega_{0}}{2}
   \big(\hat{p}^{2}+\hat{u}^{2}-1\big)
   \notag\\
   &-t_{\mathrm{kin}}\sum_{\sigma}\big(
   \hat{c}_{1\sigma}^{\dagger}\hat{c}_{2\sigma}
   +\hat{c}_{2\sigma}^{\dagger} \hat{c}_{1\sigma}\big)
   \notag\\
   &-g\hat{u}\sum_{\sigma}\big(\hat{n}_{1\sigma}-\hat{n}_{2\sigma}\big)
   \, ,
\end{align}
\end{subequations}
which is equivalent to the Hamiltonian of Eq.~\eqref{Eq:GeneralHamiltonian}
with the matrix elements
\begin{align*}
   F_{\varsigma_{p}}(z)
   &=\theta(t_{0+},z)\delta_{\varsigma_{i},1}\sqrt{2}f(z)
   \, ,\\
   \Omega_{\varsigma_{p},\varsigma_{q}}(z)
   &=\omega_{0}(\delta_{\varsigma_{p}\varsigma_{q}}
   +\alpha_{\varsigma_{p}\varsigma_{q}})/2
   \, ,\\
   M_{i\sigma,j\sigma'}^{\varsigma_{p}}(z)
   &=-g\delta_{\varsigma_{p},1}
   \delta_{\sigma\sigma'}\delta_{ij}\big(\delta_{i1}-\delta_{i2}\big)
   \, ,
\end{align*}
where we dropped the phonon mode index due to having only the relative mode.
Moreover, here $\theta$ denotes a contour-time Heaviside function and $t_{0+}$
the origin of the backward branch. The time-independent properties of this model
depend on the two dimensionless parameters
\begin{align*}
   \gamma
   &\equiv\frac{\omega_{0}}{t_{\mathrm{kin}}}
   \, ,\\
   \lambda
   &\equiv\frac{2g^2}{t_{\mathrm{kin}}\omega_{0}}
   \, ,
\end{align*}
denoting the adiabatic ratio and effective interaction.
The adiabatic ratio $\gamma$ describes the relative energy scale of
electrons and nuclei, while the effective interaction $\lambda$
is a measure of the coupling between the motions of these two constituents.


\section{Results}
\label{Sec:Results}
In the following, we present our results for the equilibrium propagators and
linear response functions. The results are for a system initially at zero
temperature in the pure two-electron $N=2$ spin singlet $S^{2}=S_{z}=0$ ground
state. This is mimicked in many-body perturbation theory with the inverse
temperature $\beta/\omega_{0}^{-1}=10^{3}$. Moreover by choosing
$G_{i\sigma,j\sigma'}(z;z')\equiv\delta_{\sigma\sigma'} G_{ij}(z;z')$
such that $N\equiv-2\imath\sum_{i}G_{ii}^{<}(t;t)=2$ for all times, we can
ensure that $S_{z}=0$. The results cover the physical parameters
$\gamma=1/2,1/4$ and $\lambda\in[0,2]$ corresponding to the weak- and
intermediate-to-strong interactions. The approximate results (H, Gd, GD) are
obtained by first solving the imaginary-time Matsubara propagators with an
imaginary-time grid, solution method and related parameters identical to the
ones used in our previous work in Ref.~\onlinecite{saekkinen-2014a}. This leads
to multiple solutions characterized by either symmetric or symmetry-broken
electron densities and nuclear displacements as shown in
Ref.~\onlinecite{saekkinen-2014a}. The former kind are the only solutions for a 
sufficiently weak interaction and are known here as the symmetric solutions,
while the latter kind arise for sufficiently strong interactions and are
referred to as asymmetric solutions. Here we mention that our approximations do 
not respect the exact transformation relating the relative and full
Hamiltonians of Ref.~\onlinecite{saekkinen-2014a}. This is seen as quantitative 
differences between some equilibirum observables, which are invariant under this 
transformation in the exact case. presented here and in
Ref.~\onlinecite{saekkinen-2014a}. In the present work, the real-time electron 
and phonon propagators are then calculated by time-propagating the Kadanoff-Baym 
equations, according to an adapted version of the algorithm~\cite{stan-2009b},
using the abovementioned Matsubara propagators either directly or indirectly
(see the linear response section) as initial values. The time-grid is uniform
with a grid spacing or time-step $\delta_{T}$ such that
$t_{\mathrm{kin}}\delta_{T}\in[0.025,0.075]$ extending from zero to the final
time $T$ chosen so that $t_{\mathrm{kin}}T=200$. The time-domain propagators are
finally Fourier transformed to arrive at their frequency-domain representations.
The Fourier transforms are calculated with a high-order quadrature formula and
unless otherwise stated by using the Hanning window function~\cite{brigham-book}.


\subsection{Equilibrium Propagators}
\label{Sec:Results:EquilibriumPropagators}
The out-of-equilibrium behavior of a system can be better understood if we first 
understand the equilibrium properties of this system. These properties are 
determined by the equilibrium electron and phonon propagators which we have
studied in Ref.~\onlinecite{saekkinen-2014a} from the perspective of time-local 
(e.g.~density matrix) and integrated-out (e.g.~total energy) observables. Here,
we further shed light on the quality of our approximations by investigating
the frequency structure of these propagators. In this section, the propagators
depend only on the relative time and our convention for evaluating Fourier 
transforms is that the first time argument is integrated over and the second
kept fixed at the initial time.


\subsubsection{Electron Propagator}
\label{Sec:ElectronPropagator}
\begin{figure*}
   \centering
   \includegraphics[height=7cm]{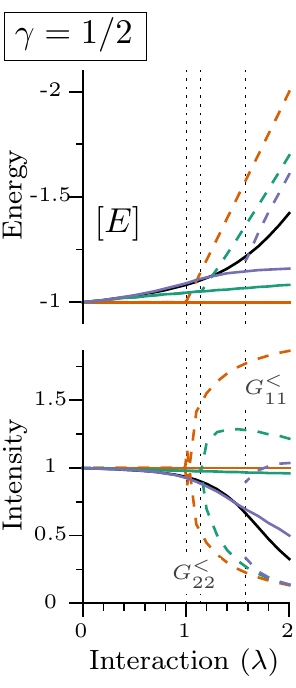}\hskip 0cm
   \includegraphics[height=7cm]{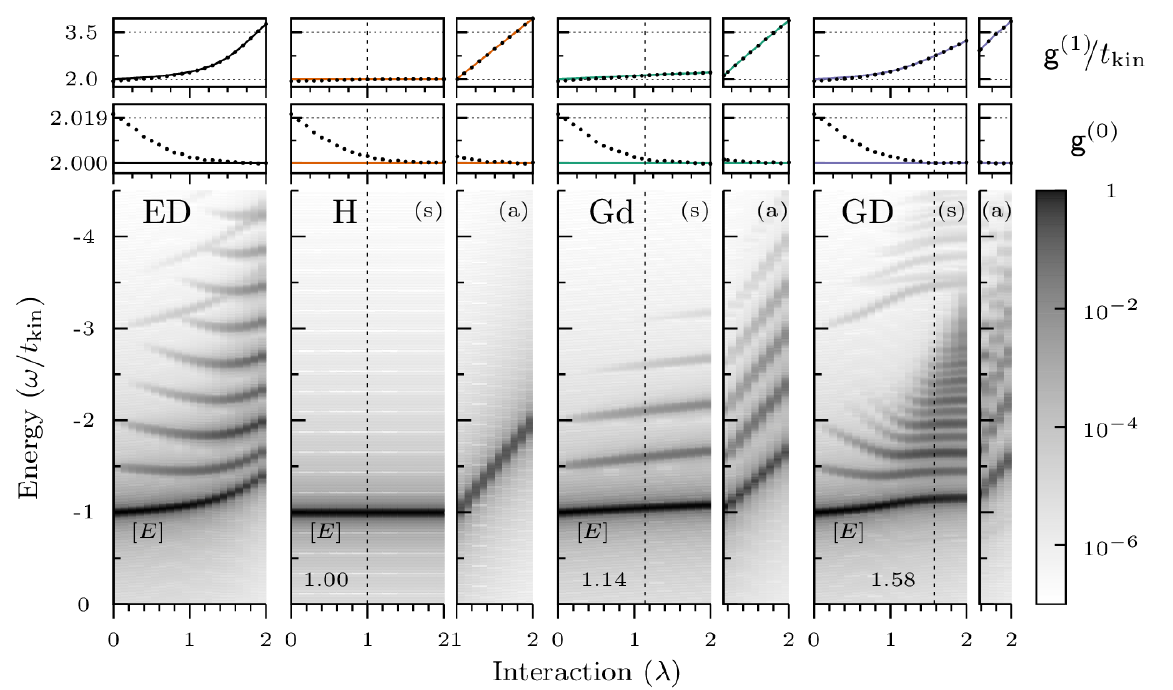}\hskip 0cm
   \includegraphics[height=7cm]{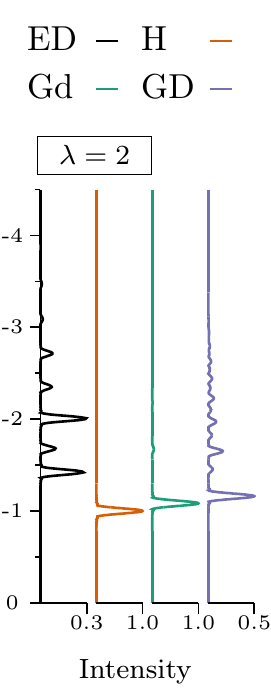}\vskip 0cm
   \includegraphics[height=7cm]{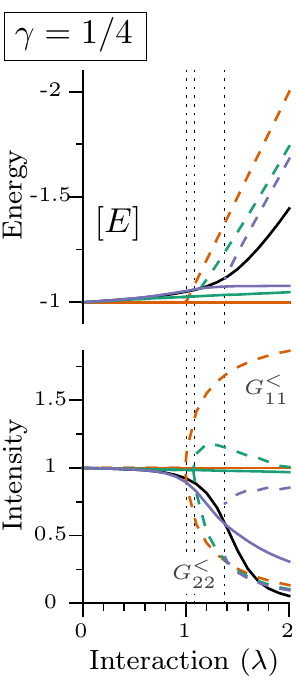}\hskip 0cm
   \includegraphics[height=7cm]{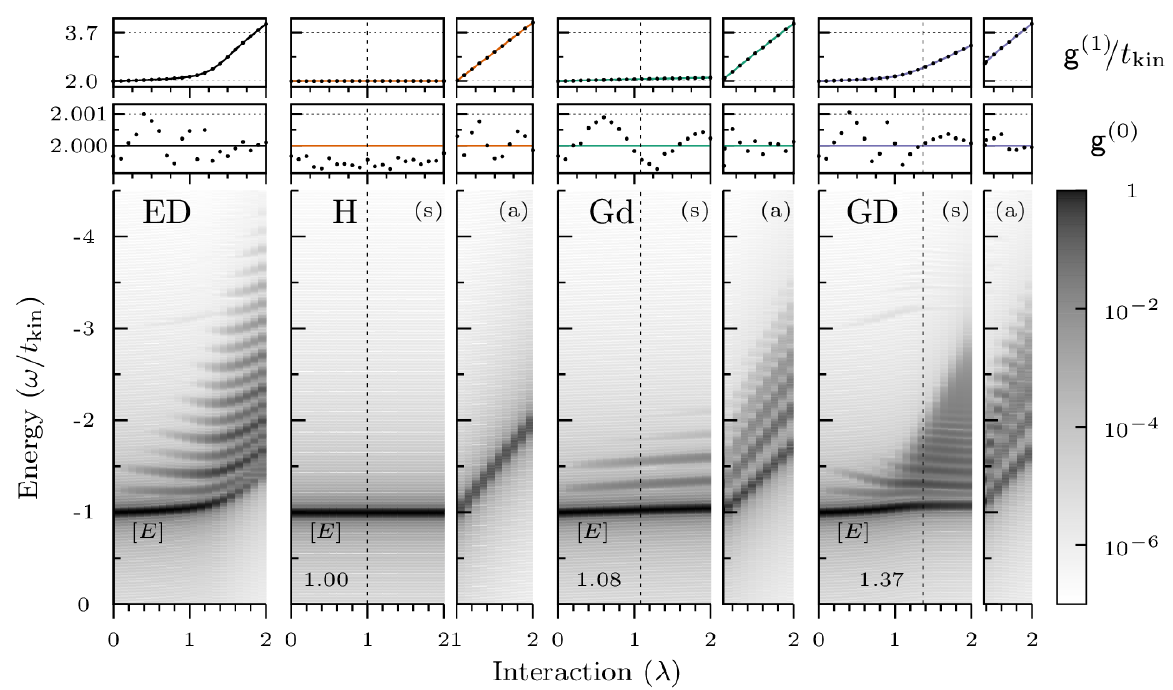}\hskip 0cm
   \includegraphics[height=7cm]{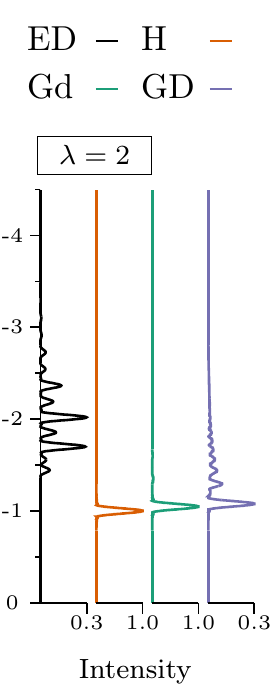}
   \caption{\label{Fig:EquilibriumElectronPropagator}
   The exact (ED) and approximate (H, Gd, GD) electron propagator as a function
   of the interaction $\lambda$ and frequency $\omega$. The top and bottom
   figures correspond to the adiabatic ratios $\gamma=1/2$ and $\gamma=1/4$, 
   respectively. The main panels in the middle show
   $\abs{\tr\M{G}^{<}(\omega)}/T$ for the symmetric (s) solutions and, starting
   from the critical interaction denoted with a vertical dashed line and its
   value $\lambda_{C}$, for the asymmetric (a) solutions. The top panels show
   the zeroth $g^{(0)}$ and first $g^{(1)}$ moments (points) of the spectra and
   the corresponding expectation values (lines) to illustrate the fulfillment of
   Eqs.~\eqref{Eq:ElectronPropagatorSumRules}. The left panels show
   the intensities and positions of the lowest energy peak of
   $2\abs{G_{11}^{<}(\omega)}/T$ and $2\abs{G_{22}^{<}(\omega)}/T$, labeled as
   $[E]$ in the main panels, for the symmetric (solid line) and asymmetric
   (dashed line) solutions. The right panel show
   $\abs{\tr\M{G}^{<}(\omega)}/T$ on a linear scale for the symmetric solutions
   at $\lambda=2$. (color online)}
\end{figure*}

The electron propagator is directly related to the photoemission
i.e.~electron removal and inverse photoemission i.e.~electron addition spectra.
This can be qualitatively seen from its zero-temperature Lehmann representation
\begin{align*}
   G^{\gtrless}_{i\sigma,j\sigma'}(\omega)
   &=\mp\imath 2\pi\sum_{n}
   f^{N\gtrless}_{n,i\sigma}f^{N\gtrless}_{n,j\sigma'}{}^{*}
   \delta(\omega\mp\Omega^{N\pm 1}_{n})
   \, ,
\end{align*}
where $\Omega^{N\pm 1}_{n}\equiv E_{n}^{N\pm 1}-E_{0}^{N}$ is the electron
addition/removal energy while
$f^{N>}_{n,i\sigma}\equiv\bra{\Psi_{n}^{N+1}}\hat{c}^{\dagger}_{i\sigma}\ket{\Psi_{0}^{N}}$
and $f^{N<}_{n,i\sigma}\equiv\bra{\Psi_{n}^{N-1}}\hat{c}_{i\sigma}\ket{\Psi_{0}^{N}}$
are the corresponding amplitudes. Here $\Psi_{n}^{N}$ and $E_{n}^{N}$ denote
the $n$th eigenstate and -energy of the $N$ electron system. The Lehmann form 
is used below to interpret the results shown in
Fig.~\ref{Fig:EquilibriumElectronPropagator} for exact diagonalization (ED)
and many-body perturbation theory (H, Gd, GD). The greater and lesser components
are related by the particle-hole symmetry
$G_{ij}^{>}(\omega)=-\big(-1)^{i-j}G_{ji}^{<}(-\omega)$ whose fulfillment is
discussed below, and therefore we only show results for the lesser component.
Let us focus first on the main panels (contour plots) to illustrate the overall
frequency structure, and start by examining the exact results. The exact spectra
develop as a function of the interaction from the singly peaked, non-interacting
spectra described by the function
\begin{align*}
   g_{ij}^{\gtrless}(\omega)
   &=\mp\imath\pi (\mp 1)^{i-j}\delta(\omega\mp t_{\mathrm{kin}})
   \, ,
\end{align*}
into spectra consisting of multiple peaks whose positions are up to an energy
shift given by the energies of the one-electron system. The one-electron
energies $E^{N=1}_{n}$ are nearly uniformly separated by the bare phonon
frequency for a weak interaction $\lambda\ll 1$. This manifests itself in the
exact spectra as emergence of the so-called phonon sideband structure which
gains intensity as the initial distribution loses intensity. In the case of a
sufficiently strong interactions the lowest energies instead consist of nearly
degenerate pairs separated by the bare phonon
frequency~\cite{ranninger-1992,alexandrov-1994}. In this case the one-elecron
system can be characterized as polaronic and is, as a first approximation in
the limit $\lambda\gg \gamma$, described by
\begin{align*}
   \ket{\psi^{\mathrm{LF}}_{k,\pm\sigma}}
   &\equiv\frac{1}{\sqrt{2}}\big(
   \hat{c}_{1\sigma}^{\dagger}\hat{X}
   \pm\hat{c}_{2\sigma}^{\dagger}\hat{X}^{\dagger}\big)\ket{0;k}
\end{align*}
where $\hat{X}\equiv\exp(-\imath g\hat{p}/\omega_{0})$ is
a shift operator, and $\ket{0;k}$ is an empty electronic state and $k$th
eigenstate of $\hat{a}^{\dagger}\hat{a}$~\cite{demello-1997}. In the same limit,
we find the two-electron ground state
\begin{align*}
   \ket{\Psi^{\mathrm{LF}}_{0}}
   &\equiv \frac{1}{\sqrt{2}}\big(
   \hat{c}_{1\uparrow}^{\dagger}\hat{c}_{1\downarrow}^{\dagger}\hat{X}^{2}
   +\hat{c}_{2\uparrow}^{\dagger}\hat{c}_{2\downarrow}^{\dagger}\hat{X}^{\dagger}{}^{2}
   \big)\ket{0;0}
   \, .
\end{align*}
which has a bipolaronic character. The removal energies and associated
amplitudes 
\begin{align*}
   \Omega_{k,l\sigma}^{1}
   &=3t_{\mathrm{kin}}\lambda/4+\omega_{0}k
   \, ,\\
   \sum_{\sigma'}\sum_{l\in\{\pm\}}
   \abs{f_{k,l\sigma';i\sigma}^{2<}}^{2}
   &=\frac{e^{-\lambda/4\gamma}}{2 k!}
   \bigg(\frac{\lambda}{4\gamma}\bigg)^{k}
   \, .
\end{align*}
then show that the spectra consist of peaks separated by the bare phonon
frequency with intensities following a Poisson
distribution~\cite{demello-1997}. The exact results shown in
Fig.~\ref{Fig:EquilibriumElectronPropagator} indicate that the initial spectra
become denser as interaction is increased such that the two lowest excitations 
approach one another faster than the third which stays roughly a bare phonon
frequency apart, especially for $\gamma=1/4$. At the same time spectral weight
is redistributed in particular to the third and higher-lying excitations.
We interpret this as a precursor of the crossover to a Poissonian disribution
which is a signature of a polaronic one-electron and bipolaronic two-electron
system. This change is accompanied by an overall shift of
the spectra to higher energies which appears smoothly as a function of the
interaction, although more rapidly around $\lambda\sim 1$ for the adiabatic
ratio $\gamma=1/4$. The shift implies that one needs more energy to either add
or remove electrons indicating that the two-electron ground state becomes more
stable. This is in agreement with the increase in the bipolaron binding
energy, see e.g.~Ref~\onlinecite{saekkinen-2014a}, and is hence associated with
the fact that the two-electron ground state becomes characterizable as
bipolaronic. In addition to these changes there is a faint signal around
$\omega/t_{\mathrm{kin}}\sim -3$ for $\gamma=1/2$ and weak interactions, which
is to be understood as the removal energy associated with the anti-bonding
state of the one-particle system. This feature is washed out for the lower
adiabatic ratio $\gamma=1/4$, in contrast to a similar feature of
the single-electron case~\cite{alexandrov-1994}.

The question is then how well the many-body approximations reproduce
the qualitative features of these spectra and thus the associated physics.
The Hartree approximation leads to spectra with peaks located at the eigenvalues
of the equilibrium Hartree equations. In the case of the symmetric solution,
the Hartree potential vanishes, and this approximation just reproduces
the non-interacting result $g_{ij}^{\gtrless}(\omega)$ for all values of
the interaction thus failing to describe the exact spectra. The asymmetric
case however displays a more complicated behavior with the propagators given by
\begin{align*}
   G_{\mathrm{H}_{a+}ij}^{\gtrless}(\omega)
   &=\imath\pi\lambda^{-1}
   \delta(\omega\mp t_{\mathrm{kin}}\lambda)
   \, ,\;i\neq j
   \, ,\\
   G_{\mathrm{H}_{a+}ii}^{\gtrless}(\omega)
   &=\mp\imath\pi
   \Big(1\pm(-1)^{i}\sqrt{1-\lambda^{-2}}\Big)
   \delta(\omega\mp t_{\mathrm{kin}}\lambda)
   \, ,
\end{align*}
where $H_{a+}$ denotes the asymmetric solution with a positive density
difference $n_{1\sigma}-n_{2\sigma}$. The asymmetric spectra emerge at
$\lambda=1$,  and contain a single peak which moves to higher energy linearly as
a function of the interaction. The particle-hole symmetry is broken along with
the reflection symmetry, however they are replaced by
$G_{\mathrm{H}_{a+}ij}^{\gtrless}(\omega)
=G_{\mathrm{H}_{a-}ji}^{\gtrless}(\omega)$ and
$G_{\mathrm{H}_{a+}ij}^{>}(\omega)
=-\big(-1)^{i-j}G_{\mathrm{H}_{a-}ji}^{<}(-\omega)$ where $H_{a-}$ denotes
the asymmetric solution with a negative relative density. These relations
represent the original symmetries under the interchange of the two degenerate
asymmetric solutions. Although the asymmetric solutions lead to spectra which
shifts to higher energies as the exact spectra do. they do not show signs of
the phonon sideband structure. The Born approximations correct this flaw and
show a clear sideband structure. The partially self-consistent Born
approximation, in the case of the symmetric solution, however shows that all
removal energies behave roughly in a similar fashion, namely they increase
monotonously and nearly linearly as a function of the interaction. The spectra
do not show signs of a peak corresponding to a removal energy associated with
the anti-bonding state of the one-particle system. This feature instead emerges 
qualitatively correctly in the fully self-consistent approximation. The fully
self-consistent approximation also improves the position of the dominant removal
energies for weak interactions by showing a stronger increase of the lowest
removal energy and a simultaneously decrease in the sideband removal energies.
Moreover, on the contrary to the monotonous behavior of the partially
self-consistent approximation, the fully self-consistent approximation shows a
signature of a stronger change in the structure of the spectrum for
$\lambda=1/4$ approximately where the exact spectrum also changes. At this
point, the fully self-consistent spectrum however becomes too dense, and does
not shift correctly to higher energies. The asymmetric solutions, once they
appear for a sufficiently strong interaction, are similar in these
approximations and differ from the asymmetric mean-field solution by the fact
that there is a related sideband structure.

The changes in the spectra from a non-interacting to a fully interacting case
should emerge in a way which respects the two lowest order sum rules for the
electron propagator
\begin{subequations}
\label{Eq:ElectronPropagatorSumRules}
\begin{align}
\label{Eq:ElectronPropagatorZerothSumRule}
   \mathsf{g}^{(0)}
   &\equiv\int\limits_{-\infty}^{\infty}\!\frac{d\omega}{2\pi\imath}\;
   \tr\M{G}^{<}(\omega)
   \notag\\
   &=N \, ,
   \\
\label{Eq:ElectronPropagatorFirstSumRule}
   \mathsf{g}^{(1)}
   &\equiv\int\limits_{-\infty}^{\infty}\!\frac{d\omega}{2\pi\imath}\;
   \omega \tr \M{G}^{<}(\omega)
   \notag\\
   &=E_{e}+E_{ep} \, ,
\end{align}
\end{subequations}
where the right-hand sides are equilibrium expectation values of the electron
number $N$, and electron $E_{e}$ and electron-phonon interaction $E_{ep}$
energies, see Ref.~\onlinecite{saekkinen-2014a}. The top panels of
Figs.~\ref{Fig:EquilibriumElectronPropagator} show that both constraints are
fulfilled up to a numerical accuracy. The numerical deviations especially for
$\gamma=1/2$ are due to choice of time discretization and frequency integration.
Moreover, we note that all frequency moments have been calculated in the present
work from spectra obtained using a rectangular window function. The first
moments, which are equal to the mean of the distribution, show that in addition
to the asymmetric cases, only the exact and symmetric fully self-consistent
spectra move appreciably to higher energies, in particular for $\gamma=1/4$.

The left panels of Fig.~\ref{Fig:EquilibriumElectronPropagator} show
the position and intensity of the lowest lying peak labeled with $[E]$, as in
Electronic, of the removal spectra. This peak is the most significant part of
the spectra in the regime of weak to intermediate interactions where the
many-body approximations are expected to be in qualitative, or even
quantitative, agreement with the exact solution. Our results show that, out of
the symmetric solutions, the Born approximations are indeed in a good agreement
with exact results in the weak coupling regime. The partially self-consistent 
approximation however deviates considerably already for intermediate
interactions $\lambda\sim 1$, while the fully self-consistent approximation
gives a reasonably good estimate of both the position and intensity up to
borderline strong interactions $\lambda\sim 1.5$. For stronger interactions,
both approximations fail to describe the shift of the position, as well as the 
decrease of the intensity correctly, although the fully self-consistent
approximation gets the latter trend considerably better. The exact position and 
intensity of this peak change more abruptly in the case of $\gamma=1/4$, and
imply that the sidebands become the most intense part of the exact spectra for
the higher interactions considered in this work. The many-body approximations do
not show sufficient loss of intensity, and therefore fail to redistribute the
spectral weight correctly to the higher energy part. The results for $\lambda=2$
shown in the right panels of Fig.~\ref{Fig:EquilibriumElectronPropagator} verify
this statement and moreover show that the approximate spectra do not bear
resemblance to the shape of the exact spectra. Lastly, the asymmetric solutions 
capture the loss of the intensity qualitatively correctly for the site with
the lower occupation but in doing so break the reflection symmetry
which leads to an increase of the intensity of the site with a higher
occupation. This is natural since it becomes favorable to remove electrons from
an already almost fully occupied site and vice versa. 

To summarize, we found that for the adiabatic ratios considered here
the Hartree, and partially and fully self-consistent Born approximations are in
a good agreement with exact results for very weak $\lambda\ll 1$, weak
$\lambda<1$, and intermediate $\lambda\sim 1$ interactions, respectively.
Moreover, the agreement between exact and approximate results improves when the
electronic and phononic energy scale become closer to one another for
$\gamma=1/2$. These observations are similar to the conclusions of our earlier
work in Ref.~\onlinecite{saekkinen-2014a} in which it was further observed that
when approaching the anti-adiabatic limit the approximate results start to again
deviate from the exact results. In particular, the comparison of the total
energies and natural occupation numbers conducted in our previous work supported
the view that the fully self-consistent approximation describes the bipolaron 
crossover partially. The present results show that as the interaction $\lambda$
is increased none of the approximate removal spectra i) move to higher energies
as in $\sim 3t_{\mathrm{kin}}\lambda/4$ nor ii) develop towards a uniformly
$\omega_{0}$-spaced distribution with a Poissonian-like envelope. The results
are consistent with our earlier findings as the sum rules are satisfied and
e.g.~$E_{e}+E_{ep}$ does show a clear significant increase in the fully
self-consistent approximation. As discussed above, points i) and ii) signal
a bipolaronic system, and their incorrect description rather suggest the
conclusion that none of the approximations describe the bipolaronic crossover
even partially. The failure to describe ii) is related to the observation that
the intensity of the lowest excitation energy does not decay fast enough as
a function of the interaction in the approximate results. This is analogous to
the insufficiently fast decaying quasi-particle spectral weight used as an
indicator of absence of the bipolaronic metal-insulator transition
in the fully self-consistent approximation~\cite{capone-2003}. Finally, our
observation on the relation between the frequency-resolved and integrated-out
quantities is similar to those obtained earlier e.g.~for the GW approximation in
the homogeneous electron gas in which self-consistent total energies were good
but the plasmon description inadequate~\cite{holm-1998}.


\subsubsection{Phonon Propagator}
\label{Sec:PhononPropagator}
\begin{figure}
   \centering
   \includegraphics[height=2.5cm]{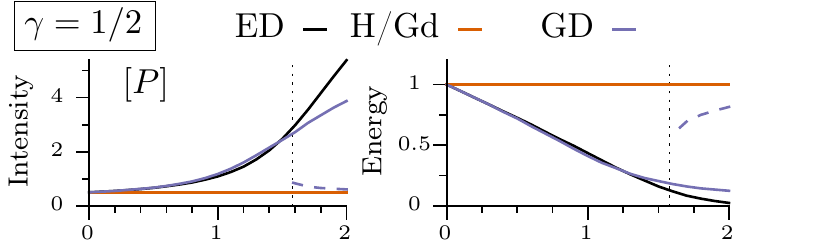}\vskip 0cm
   \includegraphics[height=6.5cm]{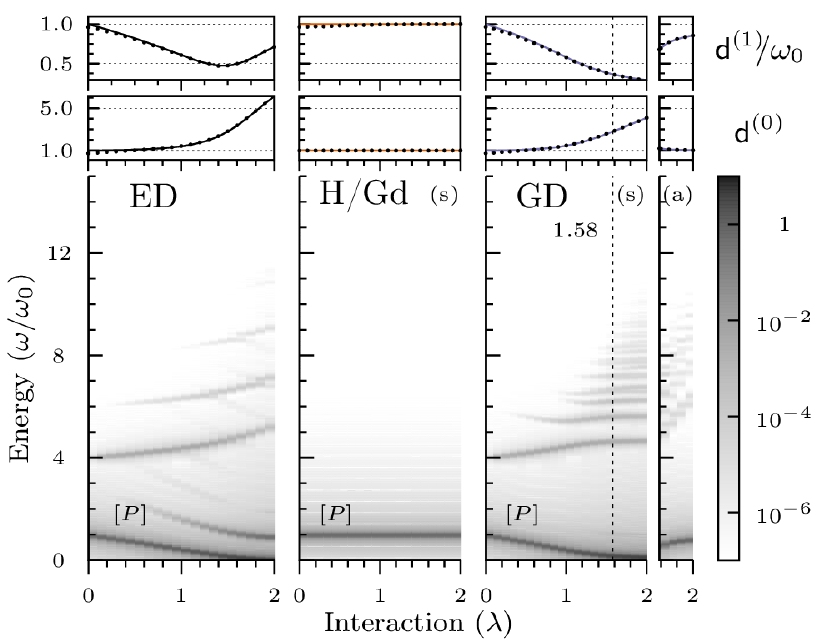}\vskip 0cm
   \includegraphics[height=2.5cm]{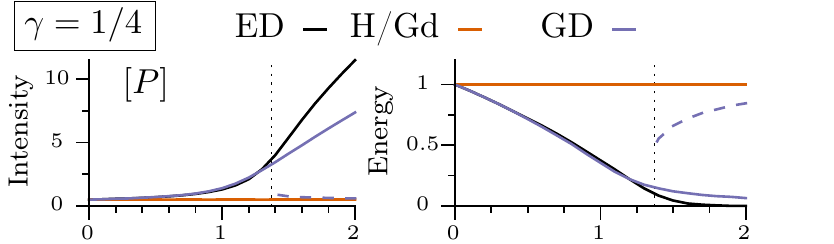}\vskip 0cm
   \includegraphics[height=6.5cm]{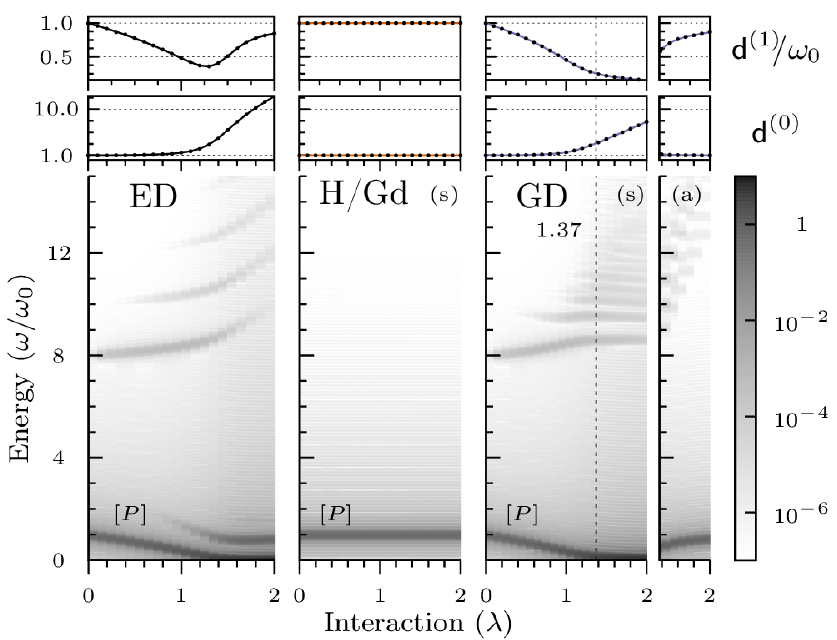}
   \caption{\label{Fig:EquilibriumPhononPropagator}
   The exact (ED) and approximate (H, Gd, GD) phonon propagator as
   a function of the interaction $\lambda$ and frequency $\omega$. The top and
   bottom figures relate to the adiabatic ratio $\gamma=1/2$ and
   $\gamma=1/4$, respectively. The contour plots show
   $\abs{D_{11}^{>}(\omega)}/T$ on logarithmic scale for the symmetric (s)
   solutions and, starting from the critical interaction marked with a vertical
   dashed line and its value $\lambda_{C}$, for the asymmetric (a) solutions.
   The middle panels show the zeroth $\mathsf{d}^{(0)}$ and first
   $\mathsf{d}^{(1)}$ moments (points) of the spectra and the corresponding
   expectation values (lines) to illustrate fulfillment of
   Eqs.~\eqref{Eq:PhononPropagatorSumRules}. The top panels show
   the intensities and positions of the lowest energy peak of
   $\abs{D_{11}^{>}(\omega)}/T$ for the symmetric (solid line) and asymmetric
   (dashed line) solutions labeled as $[P]$ in the contour plots.
   (color online)}
\end{figure}

\begin{figure}
   \centering
   \includegraphics[height=8.3cm]{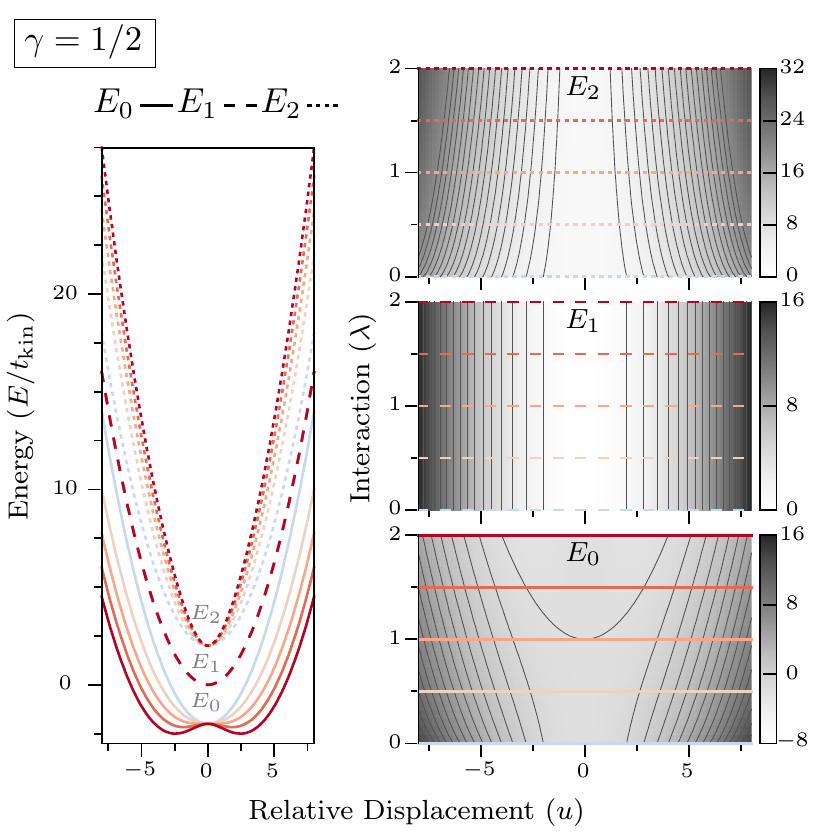}\vskip 0cm
   \includegraphics[height=8.3cm]{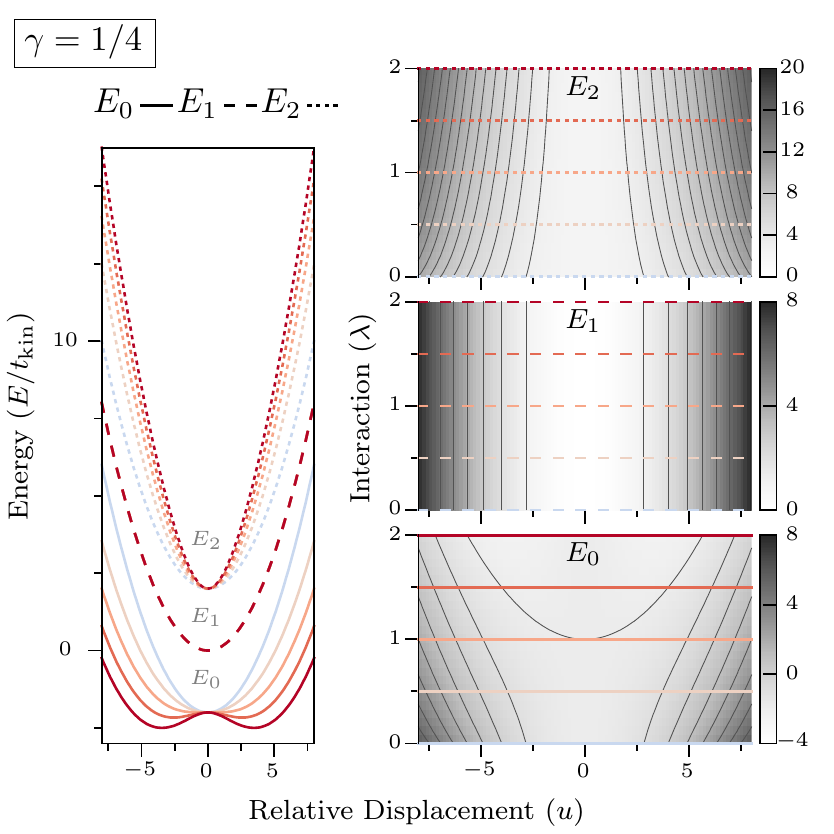}
   \caption{\label{Fig:PotentialEnergySurfaces}
   The adiabatic potential energy surfaces $E_{0}(u)$, $E_{1}(u)$ and $E_{2}(u)$
   for the three singlet eigenstates of the electronic clamped nuclei
   Hamiltonian as a function of the interaction $\lambda$ and relative
   displacement $u$, see Ref.~\onlinecite{saekkinen-2014a} for details. The top
   and bottom figures correspond to the adiabatic ratios $\gamma=1/2$ and
   $\gamma=1/4$, respectively. The left panels contain
   $\lambda=0,0.5,1.0,1.5,2.0$ cross-sections of the potential energy surfaces  
   shown as contour plots in the right panels.
   (color online)}
\end{figure}

The phonon propagator is an indicator of the properties of the nuclear system,
and relates to neutral excitations, as shown by its zero-temperature Lehmann 
representation
\begin{align*}
   D^{>}_{PQ}(\omega)
   &=D^{<}_{QP}(-\omega)
   \notag\\
   &=-\imath 2\pi\sum_{n}
   f^{N}_{n,P}f^{N}_{n,Q}{}^{*}
   \delta(\omega-\Omega^{N}_{n})
\end{align*}
where $\Omega^{N}_{n}\equiv E_{n}^{N}-E_{0}^{N}$ is a neutral excitation energy,
and $f^{N}_{n,P}\equiv\bra{\Psi_{n}^{N}}\Delta\hat{\phi}_{P}\ket{\Psi_{0}^{N}}$
the corresponding amplitude. The frequency-domain phonon propagators obtained by
means of exact diagonalization (ED) and many-body theory (H, Gd, GD) are shown
in Fig.~\ref{Fig:EquilibriumPhononPropagator}. Let us first discuss the contour
plots which illustrate the overall frequency structure of the spectra. The exact
results show that as the interaction is increased the initial, non-interacting
spectra described by
\begin{align*}
   d^{>}_{PP}(\omega)
   &= -\imath 2\pi \delta(\omega-\omega_{0})
   \, ,\\
   d^{>}_{PQ}(\omega)
   &=-2\pi(P-Q)\delta(\omega-\omega_{0})
   \, , \; P\neq Q
   \, ,
\end{align*}
where $P,Q\in\{1,2\}$ with $1$ and $2$ referring to the relative displacement
and momentum, develop into multi-peaked spectra consisting of a low and a high
energy scale. The low energy scale consists for a sufficiently strong
interaction of a single high intensity peak accompanied by a weaker peak
separated approximately by the bare phonon frequency. The high intensity peak
which is labeled with $[P]$ referring to Polaronic in the figures, develops
continuously from the initial distribution and moves rapidly towards zero energy
as a function of the interaction strength. This is true for both adiabatic
ratios with the difference that $[P]$ approaches zero more abruptly for
$\gamma=1/4$. The high energy scale, on the other hand, consists of multiple low 
intensity peaks above the first electronic excitation energy of the
non-interacting system. As the interaction is increased, these excitations move
towards higher energies and, although initially gain intensity, become
suppressed for a sufficiently strong interaction. These features can be
understood from the adiabatic potential energy surfaces defined and analyzed in
Ref.~\onlinecite{saekkinen-2014a} and shown here in
Fig.~\ref{Fig:PotentialEnergySurfaces}. This figure shows that the initially
quadratic lowest potential energy surface $E_{0}(u)$ becomes more shallow as the 
interaction is increased which is seen in
Fig.~\ref{Fig:EquilibriumPhononPropagator} as a decreasing phonon frequency.
The surface builds up a double-well structure for $\lambda>1$ which manifests
itself in the exact results as a nearly degenerate ground and first excited
state $[P]$. Moreover, as the barrier between the wells increases, the low
energy spectra approach the harmonic spectra of the isolated wells which is seen
in the exact results for $\lambda=2$ as a single peak located roughly at
the bare phonon frequency. The high-energy spectra, on the other hand, agree
with the first excited state surface $E_{1}(u)$ remaining roughly quadratic
while the surface separation $E_{1}(u)-E_{0}(u)$ increases. As discussed in
Ref.~\onlinecite{saekkinen-2014a}, in the adiabatic case $\gamma<1$ the
double-well structure is correlated with a splitting of the nuclear ground
state probability distribution and the crossover to a bipolaronic state. In
this section, we thus identify its spectral signature, that is the low energy
part consisting of the two peaks, as an indicator of a bipolaronic state.

Let us then focus on the approximate results. The Hartree and partially
self-consistent Born approximations approximate the phonon propagator with the
non-interacting propagator which does not describe the true behavior of the
interacting system discussed above. The question is then how the fully
self-consistent approximation, in which the self-energy is a single polarization
bubble, fares in this system. In order to answer this, we start with the
symmetric solution for which the contour plots of
Fig.~\ref{Fig:EquilibriumPhononPropagator} show that both energy scales of the
exact solution are reproduced for the interaction strengths considered. However,
in the low energy scale, we only observe $[P]$ and do not find a clear signature
of a peak around $\omega/\omega_{0}\sim 1$ for $\lambda\sim 2$ for the
propagation times accessed in this work. In the high energy scale, as the 
interaction is increased the fully self-consistent spectra become denser with
non-uniformly separated peaks which do not move as a whole to higher energies.
These observations are all in a disagreement with the exact results which show 
uniformly two bare phonon frequency separated peaks moving to higher energies.
This observation is however consistent with the previously discussed properties
of the approximate electron propagator for strong interactions. The asymmetric 
solution, once it is found, is observed to approach the non-interacting result 
i.e.~the lowest frequency approaches the bare phonon frequency and higher lying 
structure looses intensity as the interaction is increased. This is expected
since there is no room for particle-hole excitations in the symmetry-broken
system, and thus the polarization bubble should tend to zero when the
interaction is increased. As in the electronic case, also these spectra should 
fulfill sum rules given in terms of the zeroth and first moments by
\begin{subequations}
\label{Eq:PhononPropagatorSumRules}
\begin{align}
\label{Eq:PhononPropagatorZerothSumRule}
   \mathsf{d}^{(0)}
   &\equiv-\int\limits_{-\infty}^{\infty}\!\frac{d\omega}{2\pi\imath}\;
   \tr\M{D}^{>}(\omega)
   \notag\\
   &=\tr{\M{\Lambda}} \, ,
   \\
\label{Eq:PhononPropagatorFirstSumRule}
   \mathsf{d}^{(1)}
   &\equiv-\int\limits_{-\infty}^{\infty}\!\frac{d\omega}{2\pi\imath}\;
   \omega\tr\big(\M{\alpha}\M{D}^{>}(\omega)\big)
   \notag\\
   &=2E_{pQ}+E_{epQ}+\tr\big(\M{\alpha}\M{\Omega}^{M}\big)
   \, ,
\end{align}
\end{subequations}
where $\M{\Lambda}\equiv\imath\M{D}^{M}(0^{+})$, and $E_{pQ}$ and $E_{epQ}$
are the quantum contributions to the phonon and electron-phonon interaction
energies defined in Ref.~\onlinecite{saekkinen-2014a}. The top
panels of Fig.~\ref{Fig:EquilibriumPhononPropagator} show that these sum rules
are approximately obeyed, and therefore an important consistency relation is
satisfied. Here it is noteworthy that although there is a clear change in
the phonon energy, see the zeroth frequency moment, in the exact and fully
self-consistent solutions, only the former displays a clear kink
at $\lambda\sim 1.3-1.5$ in the first frequency moment. Finally, the top panels
of Fig.~\ref{Fig:EquilibriumPhononPropagator} highlight the lowest excitation
energy labeled with $[P]$ which is the dominant part of the spectra. The exact 
results show that this peak approaches zero energy, but never actually reaches
it, and gains intensity as a function of the interaction. This is in contrast to
the non-interacting propagator in which this sole feature remains at the bare
phonon frequency. The self-consistent Born approximation however captures both 
effects reasonably accurately up to $\lambda\sim 1.5$ and gives a qualitatively 
similar trend even beyond it for the interactions considered here.

To summarize, we found that the non-interacting propagator used in the Hartree
and partially self-consistent Born approximations is an adequate approximation
for the interacting phonon propagator only for very weak $\lambda \ll 1$ 
interactions. The fully self-consistent Born approximation, on the other hand, 
captures the dominant low energy peak well up to borderline strong
$\lambda\sim 1.5$ interactions and is therefore in a good agreement with the
exact results in this range of interactions. It however does not describe the
second low energy excitation at $\omega_{0}$, and therefore reproduces
qualitatively only one of the signatures of a bipolaronic state observable in
the phonon propagator. Finally, we remark that the absence of a peak at the bare 
phonon frequency for strong interactions is a likely factor for the observed
too dense frequency structure of the electron propagator.


\subsection{Linear Response Functions}
The dynamics of a system of electrons and nuclei is in many cases dominantly
determined by a linear response function provided that the system is perturbed
sufficiently weakly. Many spectroscopic methods essentially rely on measuring
these functions which makes them important for understanding experiments.
Here we investigate in particular the first order density-density response
function of our model system.


\subsubsection{Method}
\label{Sec:ResultsMethod}
Let us begin by explaining how we in practice calculate linear response
functions in our time-dependent formalism. This is a prerequisite for
understanding when it is reasonable to do linear response by time-propagation.
The density-density response function is calculated by perturbing the system
with the time-dependent potential
\begin{align}
\label{Eq:DeltaPotential}
   \op{V}(t)
   &=\delta(t)\sum_{i\sigma}v_{i}\op{n}_{i\sigma}
   \, ,
\end{align}
where $v_{i}$ is the magnitude of the perturbation, and $\delta$ is the Dirac
delta function. Then we record the resulting spin-summed density
\begin{align*}
   n_{i}(t)
   &=-2\imath G_{ii}^{<}(t;t)
   \, ,
\end{align*}
which in the linear response regime satisfies
\begin{align}
\label{Eq:LinearResponseDensity}
   \delta n_{i}(t)
   &\equiv n_{i}(t)-n_{i}^{(0)}(t)
   \notag\\
   &=\sum_{j}\chi^{R}_{ij}(t)v_{j}
   +\mathcal{O}(v^{2})
   \, ,
\end{align}
where $v$ is the norm of a vector with $v_{i}$ as its components,
$n_{i}^{(0)}(t)$ is the density of the unperturbed system, and
$\chi^{R}_{ij}(t)\equiv\sum_{\sigma\sigma'}\chi_{i\sigma,j\sigma'}^{R}(t)$
the retarded component of the first order density-density response function.
The response function is then given by
\begin{align}
\label{Eq:RetardedDensityResponseFunction}
   \chi_{ij}^{R}(t)
   &\equiv\frac{\partial n_{i}(t)}{\partial v_{j}}\bigg|_{v=0}
   \, ,
\end{align}
which we in practice evaluate by using the difference quotient
$(n_{i}(t)-n_{i}^{(0)}(t))/v_{j}$ with $v_{j}$ sufficiently small
and $v_{k}=0$ for $k\neq j$. Lastly, it is important to
understand that applying the delta function potential amounts to choosing a new
initial state after which the time-evolution is induced by the unperturbed
Hamiltonian. In exact diagonalization, this is achieved by preparing the new
initial state
\begin{align}
\label{Eq:NewExactInitialState}
   \ket{\tilde{\Psi}_{0}^{N}}
   &=e^{-\imath\sum_{i\sigma}v_{i}\op{n}_{i\sigma}}\ket{\Psi_{0}^{N}}
   \, ,
\end{align}
where $\ket{\Psi_{0}^{N}}$ is the $N$ electron ground state, and which is
subsequently propagated in the absence of the perturbation. In the Kadanoff-Baym 
equations, on the other hand, the same is achieved by choosing the new initial 
electron propagators
\begin{align}
\label{Eq:NewApproximateInitialState}
   G_{ij}^{\gtrless}(0;0)
   &=e^{-\imath (v_{i}-v_{j})} G_{ij}^{\mathrm{M}}(0^{\pm})
   \, ,\\
   G_{ij}^{\rceil}(0;\tau)
   &=e^{-\imath v_{i}} G_{ij}^{M}(-\tau)
   \, ,
\end{align}
where $G_{ij}^{\mathrm{M}}(\tau)$ is the solution to the equilibrium Dyson
equation. The electron and phonon propagators are then obtained by
time-propagating the unperturbed Kadanoff-Baym equations.


\subsubsection{Stability}
\label{Sec:Stability}
The method described above is expected to work if the perturbation expansion
of Eq.~\eqref{Eq:LinearResponseDensity} is valid for the time scales
of interest. It can however be that a possibly finite time-scale in which
the expansion is good cannot be extended to cover the entire time-scale of
interest. This can signal e.g.~an unbounded linear response function.
In the following, we show that this is the case for the Hartree approximation,
and subsequently investigate whether or not the Born approximations show
a similar behavior.

The Hartree Eqs.~\eqref{Eq:HartreeEquations} are a closed set of ordinary
differential equations for the phonon field expectation value $\phi_{P}(t)$ and
the electron propagator $G_{ij}^{<}(t;t)$. In the two-site, two-electron
Holstein model these equations can be rewritten as
\begin{subequations}
\label{Eq:HartreeEquationsForDimer}
\begin{align}
   \dot{n}
   &=4t_{\mathrm{kin}}\Gamma_{2}\, ,\\
   \dot{\Gamma}_{1}
   &=-2gu\Gamma_{2}\, ,\\
   \dot{\Gamma}_{2}
   &=-t_{\mathrm{kin}}n+2gu\Gamma_{1}\, ,\\
   \dot{u}
   &=\omega_{0}p\, ,\\
   \dot{p}
   &=-\omega_{0}u+2gn\, ,
\end{align}
\end{subequations}
where we have suppressed the time arguments and the overhead dot denotes
the time-derivative. Moreover, $n\equiv n_{1\sigma}-n_{2\sigma}$,
$u\equiv (u_{1}-u_{2})/\sqrt{2}$, and $p\equiv (p_{1}-p_{2})/\sqrt{2}$ are
the relative spin density, displacement, and momentum, while $\Gamma_{1}$ and
$\Gamma_{2}$ are the real and imaginary parts of the density matrix element
$\gamma_{12}\equiv-\imath G_{12}^{<}$, respectively. As shown explicitly in
App.~\ref{App:HartreeDensityResponse}, the density-density response function is
the solution to the corresponding linearized equations of motion. If the 
linearization is performed with respect to an equilibrium solution which is a
stable fixed-point, in the sense of Lyapunov~\cite{hirsch-book,perko-book,
jordan-book}, of the original equations then the eigenvalues of the resulting 
Jacobian matrix have non-positive real parts~\cite{hirsch-book}. Moreover, if
there are no repeated zero eigenvalues, then the zero solution of the linearized 
system is stable and furthermore any solution is bounded~\cite{jordan-book}. In 
particular, the density response function is then bounded, that is $\exists$
$M>0$ independent of $t$ such that $\abs{\chi_{ij}^{R}(t)}\leq M$ for all
$t\geq 0$.  In order understand when this is the case, we investigate below the 
stability of the fixed-points of the Hartree equations. The fixed-points whose 
stability is to be studied are just the symmetric
\begin{subequations}
\label{Eq:HartreeFixedPoints}
\begin{align}
\label{Eq:SymmetricHartreeFixedPoint}
   n_{s}
   &=0
   \, ,\notag\\
   \Gamma_{s,2}
   &=0
   \, ,\notag\\
   u_{s}
   &=0
   \, ,\notag\\
   p_{s}
   &= 0
   \, ,
\end{align}
and asymmetric
\begin{align}
\label{Eq:AsymmetricHartreeFixedPoint}
   n_{a\pm}
   &=\pm\sqrt{1-\lambda^{-2}}
   \, ,\notag\\
   \Gamma_{a\pm,2}
   &=0
   \, ,\notag\\
   u_{a\pm}
   &=2gn_{a\pm}/\omega_{0}
   \, ,\notag\\
   p_{a\pm}
   &=0
   \, .
\end{align}
\end{subequations}
solutions of the equilibrium Hartree equations of
Eqs.~\eqref{Eq:EquilibriumHartreeEquations} derived in
Ref.~\onlinecite{saekkinen-2014a}. These equations are subject to two constants
of motion as both the eigenvalues of the reduced density matrix, which are
either one or zero, and the total energy are conserved and give the constraints
\begin{align}
   1
   &=n^{2}
   +4\big(\Gamma_{1}^{2}+\Gamma_{2}^{2}\big)
   \, ,\notag\\
\label{Eq:HartreeTotalEnergy}
   E
   &=\frac{\omega_{0}}{2}\big(p^{2}+u^{2}-1\big)
   -4t_{\mathrm{kin}}\Gamma_{1}
   -2g nu
   \, ,
\end{align}
respectively. Then by following~\cite{aguiar-1991,itin-2010}, and motivated by
the first constraint, we introduce the coordinates
\begin{align*}
   z
   &=2\Gamma_{1}
   \, ,\\
   n
   &=\sqrt{1-z^{2}}\cos(\theta)
   \, ,\\
   \Gamma_{2}
   &=\sqrt{1-z^{2}}\sin(\theta)/2
   \, ,
\end{align*}
which represent cross-sections of the unit sphere with a plane.
The transformed equations of motion
\begin{align*}
   \dot{\theta}
   &=-2t_{\mathrm{kin}}+2g\frac{uz\cos(\theta)}{\sqrt{1-z^{2}}}
   \, ,\\
   \dot{z}
   &=-2gu\sqrt{1-z^{2}}\sin(\theta)
   \, ,\\
   \dot{u}
   &=\omega_{0}p
   \, ,\\
   \dot{p}
   &=-\omega_{0}u+2g\sqrt{1-z^{2}}\cos(\theta)
   \, ,
\end{align*}
and the total energy
\begin{align*}
   E
   &=\frac{\omega_{0}}{2}\big(p^{2}+u^{2}-1\big)
   -2t_{\mathrm{kin}}z
   -2gu\sqrt{1-z^{2}}\cos(\theta)
   \, ,
\end{align*}
then correspond to a Hamiltonian system with canonical conjugate
variables $(\theta,z)$ and $(u,p)$. The canonical transformation
\begin{align*}
   q_{1}
   &=-\sqrt{2(1-z)}\sin(\theta)
   \, ,\\
   p_{1}
   &=\sqrt{2(1-z)}\cos(\theta)
   \, ,\\
   q_{2}
   &=-p
   \, ,\\
   p_{2}
   &=u
   \, ,
\end{align*}
transforms this system into two non-linearly coupled oscillators described by
\begin{align*}
   E
   &=\frac{\omega_{0}}{2}\big(p_{2}^{2}+q_{2}^{2}\big)
   +t_{\mathrm{kin}}\big(p_{1}^{2}+q_{1}^{2}\big)
   \notag\\
   &-2gp_{2}p_{1}\sqrt{1-\big(p_{1}^{2}+q_{1}^{2}\big)/4}
   \, ,
\end{align*}
where we dropped a constant energy shift. This system of equations is a special
case of the Hamiltonian system studied in~\cite{aguiar-1991} which arises
from the semi-classical equations of the Dicke model~\cite{aguiar-1991,
bakemeier-2013}. Let us then denote $x_{1}\equiv q_{1}$, $x_{2}\equiv q_{2}$,
$x_{3}\equiv p_{1}$, and $x_{4}\equiv p_{2}$. The fixed points of this system
are just related by coordinate transforms to the fixed points of
Eqs.~\eqref{Eq:HartreeFixedPoints}. At the symmetric fixed-point, we find
the Hessian matrix 
\begin{align*}
   \M{\nabla\nabla}E_{s}
   \equiv
   \begin{pmatrix}
      2t_{\mathrm{kin}} & 0 & 0 & 0 \\
      0 & \omega_{0} & 0 & 0 \\
      0 & 0 & 2t_{\mathrm{kin}} & -2g \\
      0 & 0 & -2g & \omega_{0} 
   \end{pmatrix}
\end{align*}
which is positive-definite for $\lambda<1$ and indefinite for $\lambda>1$,
while at the asymmetric fixed-points, the Hessian matrix
\begin{align*}
   \M{\nabla\nabla} E_{a\pm}
   &\equiv
   \begin{pmatrix}
      t_{\mathrm{kin}}(1+\lambda) &
      0 &
      0 &
      0 \\
      0 &
      \omega_{0} &
      0 &
      0 \\
      0 &
      0 &
      \frac{4t_{\mathrm{kin}}\lambda}{1+\lambda^{-1}} &
      -\frac{2\sqrt{2}g\lambda^{-1}}{\sqrt{1+\lambda^{-1}}} \\
      0 &
      0 &
      -\frac{2\sqrt{2}g\lambda^{-1}}{\sqrt{1+\lambda^{-1}}} &
      \omega_{0}
   \end{pmatrix}
\end{align*}
is positive-definite for $\lambda>1$. The symmetric equilibrium and
asymmetric equilibria are then due to the Lagrange-Dirichlet
theorem~\cite{arnold-book,krechetnikov-2007} stable for $\lambda<1$ and
$\lambda>1$, respectively. Moreover, since $\det(\M{\nabla}\M{\nabla}E_{s})<0$ 
for $\lambda>1$ also $\det(\M{J}\M{\nabla}\M{\nabla}E_{s})<0$, where $\M{J}$ is 
the standard symplectic matrix~\cite{arnold-book}, and therefore the Jacobian 
matrix of the linearized Hamilton's equations has an eigenvalue with a negative 
real-part. This implies that there also exists an eigenvalue with a positive 
real part which means that the equations are linearly and nonlinearly
unstable~\cite{jordan-book}. The symmetric equilibrium is therefore unstable for 
$\lambda>1$. The zero solution losing its stability while two new stable 
equilibria arise is a standard bifurcation known as the supercritical pitchfork
bifurcatioñ~\cite{kuznetsov-book}. The stability together with the fact that
the Hessian matrices do not have zero eigenvalues for $\lambda\neq 1$ implies
that the response functions obtained for $\lambda<1$ and $\lambda>1$ using
the symmetric equilibrium and asymmetric equilibria are bounded functions.
The linear instability of the symmetric solution for $\lambda>1$ leads,
on the other hand, to an unbounded response function as shown in
App.~\ref{App:HartreeDensityResponse}.

This answers the question when it is in this context appropriate to do linear
response properties at the mean-field level, but does not resolve this
issue for the correlated approximations. In this case, one cannot recast
the equations as a set of ordinary differential equations, but instead must
consider the full two-time integro-differential equations which have
non-linear integral kernels. As we are not aware of stability theory for such
dynamical systems and it would go beyond the scope of the present work, we only
resort to a working measure which is in the spirit of the stability of
the equilibrium solutions. The working measure chosen here is a practical one:
we introduce the norm
\begin{align}
\label{Eq:WorkingMeasureStability}
   \norm{\delta n}_{\infty}
   &\equiv\max_{t\in[0,T]}\abs{n(t)-n(0)}
   \, ,
\end{align}
compare it to the magnitude of the perturbation $v$, and if they remain
in the same order of magnitude, we suggest that the equilibrium is stable.
We emphasize that this measure is not equal to the stability even in the case of
ordinary differential equations, but does give a practical estimate whether or
not a linear response calculation makes sense for the time scales accessed
in this work.

\begin{figure}
   \centering
   \includegraphics[height=8cm]{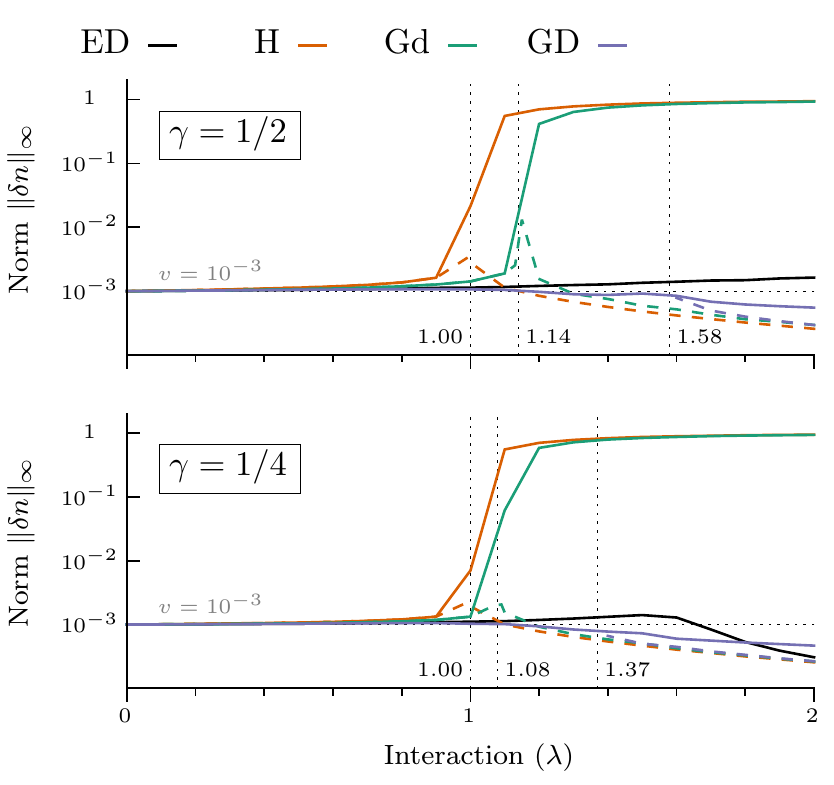}
   \caption{\label{Fig:TdDensityKick}
   The norm $\norm{\delta n}_{\infty}$ of Eq.~\eqref{Eq:WorkingMeasureStability}
   as a function of the interaction $\lambda$ for the perturbation of
   Eq.~\eqref{Eq:DeltaPotential} with $v_{1}=10^{-3},v_{2}=0$. The top and
   bottom panels correspond to the adiabatic ratios $\gamma=1/2$ and
   $\gamma=1/4$, respectively. The solid and dashed lines correspond to the
   symmetric and asymmetric solutions, respectively. The vertical dashed lines
   and the associated values $\lambda_{C}$ denote the critical interactions
   which are ordered according to to H, Gd, and GD from left to right.
   (color online)}
\end{figure}

Having said this, Fig.~\ref{Fig:TdDensityKick} shows this norm for exact 
diagonalization (ED) and many-body theory (H, Gd, GD). The results are obtained 
either by starting from a symmetric, or asymmetric equilibrium solution.
The exact results stay always in the same magnitude as the perturbing potential
$v=10^{-3}$ which is expected since the time-dependent Schr\"{o}dinger equation
is linear. The other extreme is the symmetric mean-field solution for which
the norm suddenly, although continuously as a function of the interaction,
approaches one when the interaction exceeds the corresponding critical value
$\lambda=1$. At the same time, this norm remains in the same order of magnitude
as the perturbation for the asymmetric mean-field solution. This is consistent
with the stability analysis presented above illustrates that our working measure 
agrees in this case with Lyapunov stability. These cases give us some confidence
in looking at the symmetric and asymmetric solutions of the Born approximations.
The results for the asymmetric cases of these approximations indicate that they
behave  qualitatively similar to the asymmetric equilibria of the mean-field
suggesting that they are stable. This qualitative agreement remains true for
the symmetric equilibria of the partially self-consistent case in which
the norms approach one when passing the critical value of interaction. The norm
of the symmetric solution of the fully self-consistent approximation however
remains in the same order of magnitude as the perturbation. We note that
the declining norm for $\gamma=1/4$ and higher interactions is in the exact case
related to the fact that the propagation length is shorter than the period of
the lowest excitation of the system.

These observations together with the interpretation given to our working measure
of stability then suggest that the partially self-consistent approximation has
the same qualitative stability properties as the mean-field. On the other hand,
both equilibrium solutions of the fully self-consistent Born approximation
are observed to be stable in the sense of our working measure. This means that,
on the contrary to the Hartree and partially self-consistent Born approximation,
it is possible to do linear response with respect to the symmetric equilibrium
of the fully self-consistent Born approximation for the time-scales addressed
here.


\subsubsection{Bethe-Salpeter equation}
\begin{figure}
   \centering
   \includegraphics[scale=0.7]{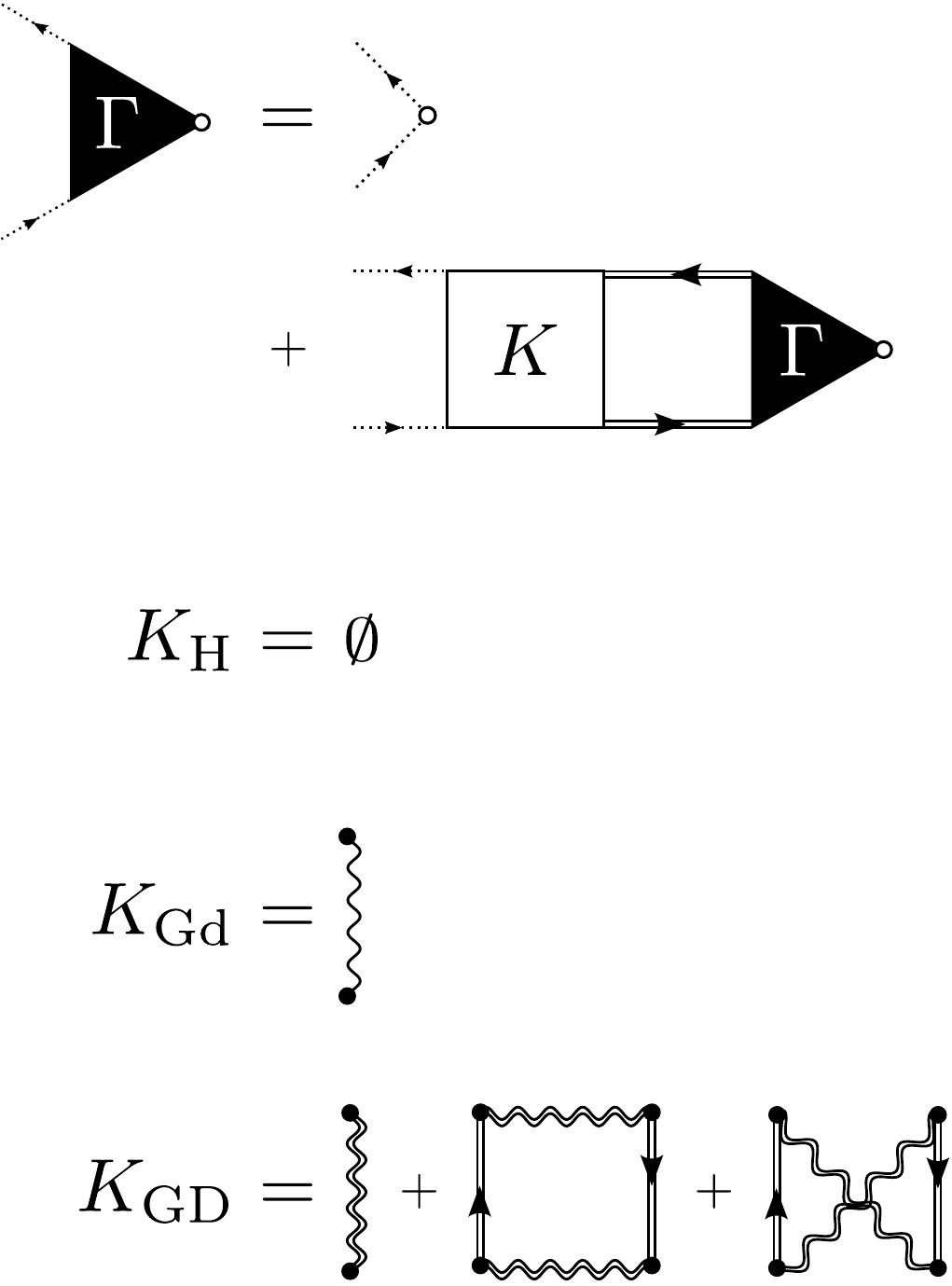}
   \caption{\label{Fig:bse-kernels}
   The Bethe-Salpeter equation for the irreducible vertex function
   (top) and functional forms of the four-point kernels (bottom) in the Hartree 
   (H), and the partially (Gd) and fully (GD) self-consistent Born
   approximations. A line with an arrow indicates a dressed electron propagator, 
   while single and two-fold wiggly lines represent bare and dressed phonon 
   propagators, respectively. An open circle represents a connection for a
   phonon propagator and a closed circle (kernel) or a dashed line (BSE) a 
   connection for an electron propagator.}
\end{figure}

The Bethe-Salpeter equation is the standard approach to calculate linear
response functions in many-body perturbation theory~\cite{onida-2002}. Here,
we discuss the connection between the many-body approximations used in this
frequency-domain approach and in the time-dependent approach applied in the
present  work. We start by noting that the density response function of
Eq.~\eqref{Eq:RetardedDensityResponseFunction} is the retarded
$\chi_{ij}^{R}(t)=\sum_{\sigma\sigma'}
\chi^{R}_{i\sigma i\sigma,j\sigma' j\sigma'}(t;0)$
component of the generalized, contour-time response function
\begin{align}
\label{Eq:GeneralizedResponseFunction}
   \chi_{ij,kl}(z;z')
   \equiv\frac{1}{\imath\Z}\Tr\Bigg[\toop\bigg\{
   e^{-\imath\int_{C}\;d\bar{z}\op{H}(\bar{z})}
   \Delta\hat{\gamma}_{ij}(z)\Delta\hat{\gamma}_{kl}(z')\bigg\}\Bigg]
   \, ,
\end{align}
where $\hat{\gamma}_{ij}\equiv\hat{c}_{j}^{\dagger}\hat{c}_{i}$ is the one-body
reduced density matrix operator and $\Delta\hat{\gamma}_{ij}$ the corresponding
fluctuation operator. Note that we switched here to collective indices
containing both spatial and spin degrees of freedom. In the standard approach,
the generalized response function satisfies the equation
\begin{subequations}
\label{Eq:DysonEquationDensityResponse}
\begin{align}
   \chi_{ij,kl}(z;z')
   &=P_{ij,kl}(z;z')
   +\sum_{PQ}\sum_{rstu}\int\limits_{C}\!d\bar{z}d\ubar{z}\;
   P_{ij,rs}(z;\bar{z})
   \notag\\
   &\times
   M_{sr}^{P}(\bar{z})d_{PQ}(\bar{z};\ubar{z})M_{tu}^{Q}(\ubar{z})
   \chi_{ut,kl}(\ubar{z};z')
   \, ,\\
\label{Eq:polarizability}
   P_{ij,kl}(z;z')
   &=-\imath\sum_{pq}\int\limits_{C}\!d\bar{z}d\ubar{z}\;
   G_{ip}(z;\bar{z})
   \notag\\
   &\times G_{qj}(\ubar{z};z)\Gamma_{pq;kl}(\bar{z},\ubar{z};z')
   \, ,
\end{align}
\end{subequations}
where $P_{ij,kl}(z;z')$ is the irreducible polarizability defined in terms of
the irreducible vertex function $\Gamma_{ij,kl}(z,z';z'')$. These equations are
valid for any many-body approximation which includes the mean-field, Hartree
term while beyond mean-field effects are incorporated into the irreducible
vertex function. This function satisfies the Bethe-Salpeter
equation~\cite{stefanucci-book}
\begin{align*}
   \Gamma_{ij,kl}(z,z';z'')
   &=\delta_{il}\delta_{jk}\delta(z,z')\delta(z',z'')
   \notag\\
   &+\sum_{pqrs}\int\limits_{C}\!d\bar{z}d\bar{z}'d\ubar{z}d\ubar{z}'\;
   K_{iq,pj}(z,\bar{z};\bar{z}',z')
   \notag\\
   &\times G_{pr}(\bar{z}';\ubar{z})G_{sq}(\ubar{z}';\bar{z})
   \Gamma_{rs,kl}(\ubar{z},\ubar{z}';z'')
   \, ,
\end{align*}
where the four-point integral kernel is defined as
\begin{align}
\label{Eq:FourPointKernel}
   K_{ij,kl}(z,z';\bar{z},\bar{z}')
   &\equiv\frac{\delta\Sigma_{\mathrm{xc},il}(z;\bar{z}')}
   {\delta G_{kj}(\bar{z};z')}
   \, ,
\end{align}
with the subscript xc denoting the exchange-correlation self-energy. The 
diagrammatic form of this equation is shown in the top panel of
Fig.~\ref{Fig:bse-kernels}. It has been shown~\cite{kwong-2000} that a density 
response function obtained by time-propagation of the Kadanoff-Baym equations
with a self-energy $\Sigma$ is equivalent to a solution of
Eqs.̃\eqref{Eq:DysonEquationDensityResponse} with the vertex satisfying the
Bethe-Salpeter equation with a four-point kernel of
Eq.~\eqref{Eq:FourPointKernel}. Thus by calculating the response function via
time-propagation using Eqs.~\eqref{Eq:LinearResponseDensity}
and~\eqref{Eq:RetardedDensityResponseFunction} we arrive by current standards
at a high-level solution of the Bethe-Salpeter equation.

Figure.~\ref{Fig:bse-kernels} shows diagrammatically the approximate four-point 
kernels related to the self-energy approximations used in the present work. In
the Hartree approximation, the irreducible vertex function is the bare vertex,
and the response function is hence the sum of all bubble diagrams. The partially 
self-consistent Born approximation leads to a vertex function which consists of
all the ladder diagrams and to the bubbles-and-ladders series for the response 
function. In addition to such terms, the fully self-consistent Born
approximation contains also two second order kernel diagrams in terms of the
phonon propagators. These higher-order terms are not routinely considered in the
zero-frequency~\cite{freericks-1993,freericks-1994} nor fully
frequency-dependent cases~\cite{onida-2002}. Lastly, we note that despite of the
sophistication of these approximations there are in general no guarantees of
their superiority over the conventional approximations. It has also been shown,
that similar approximations in the purely electronic case, can lead to undesired
features like non-positivity~\cite{uimonen-2015}.


\subsubsection{Density-Density Response Function}
\label{Sec:DensityDensityResponseFunction}
\begin{figure}
   \centering
   \includegraphics[height=9.5cm]{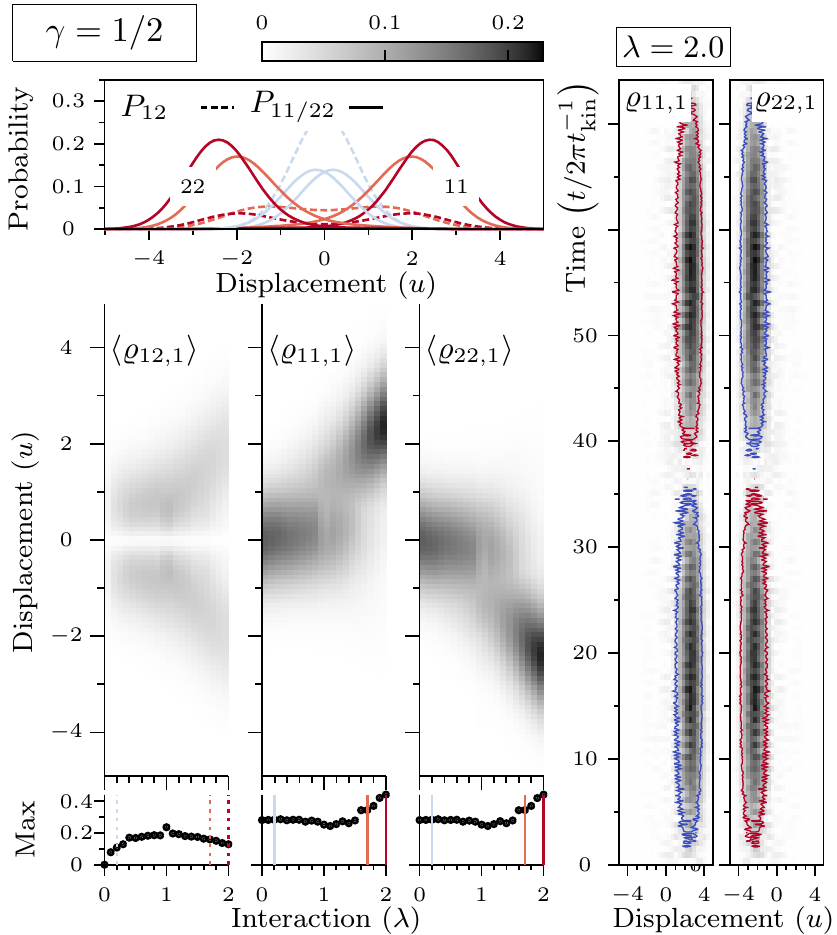}\vskip 0cm
   \includegraphics[height=9.5cm]{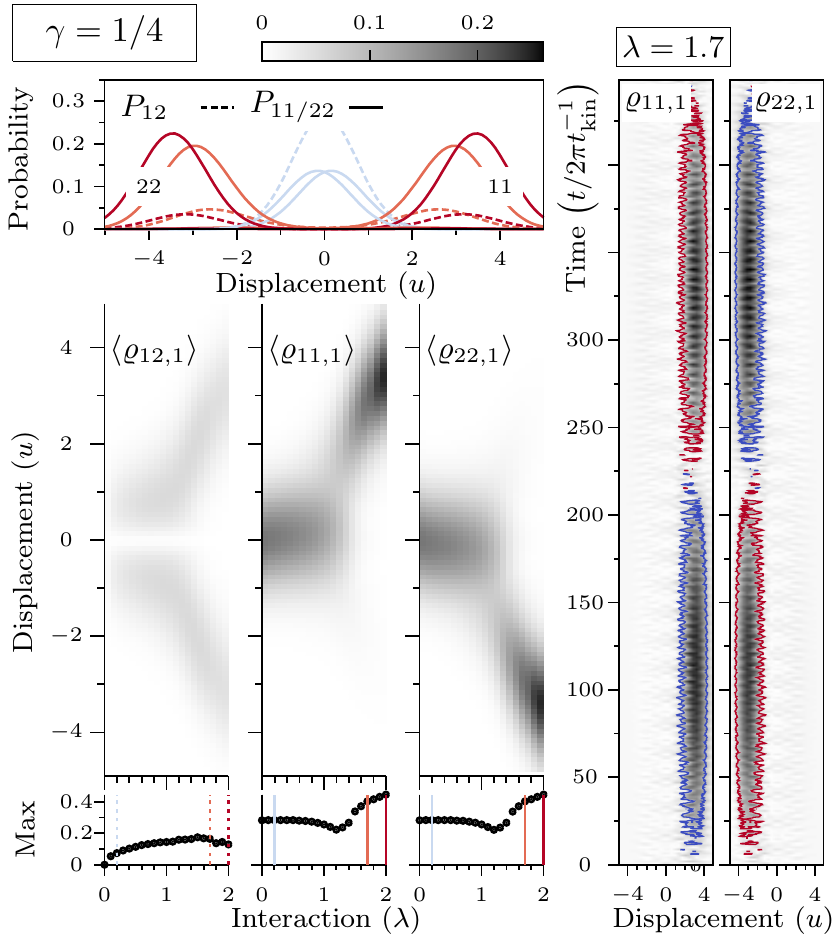}
   \caption{\label{Fig:TimeDependentJointProbability}
   The exact (ED) joint-probabilities $P_{ij}(u;0)$ and response functions
   $\varrho_{ik,j}(u;t)$ shown for the adiabatic ratios $\gamma=1/2$ and
   $\gamma=1/4$ in the top and bottom figures, respectively. In the top panels,
   we show $P_{ij}(u;0)$ as a function of the displacement $u$ for the
   interactions $\lambda=0.2,1.7,2.0$ whose color code is shown in the bottom
   panels. The middle panels contain $\langle\varrho_{ij,1}\rangle$ as a
   function of the displacement and interaction, and the bottom panels display
   $\max{u,t\in[0,T]}\abs{\varrho_{ij,1}(u;t)}$ as a function of the
   interaction. In right panels, we show $\varrho_{ii,1}(u;t)$ as a function of
   the displacement and time $t$ for the interactions $\lambda=2.0,1.7$ with
   regions closed by red (blue) contour lines being positive (negative) values.
   (color online)}
\end{figure}

We have presented a stability analysis in order to understand when it makes
sense to calculate linear response properties in the context of the present
work. Additionally, we have discussed the diagrammatic meaning of the response 
function obtained in this manner. Here we focus on the numerical results, that
is analyzing the density-density response function obtained by time-propagation
of the Kadanoff-Baym equations. In order to carry-out this analysis, we start by
considering the exact response function which has the frequency-domain Lehmann
representation
\begin{align*}
   \chi^{R}_{ij}(\omega)
   &=\sum_{n}\bigg(\frac{h_{ni}^{N}h_{nj}^{N}}
   {\omega-\Omega_{n}^{N}+\imath\eta}
   -\frac{h_{ni}^{N}h_{nj}^{N}}
   {\omega+\Omega_{n}^{N}+\imath\eta}\bigg)
   \, ,
\end{align*}
where $\Omega_{n}^{N}\equiv E_{n}^{N}-E_{0}^{N}$ is a neutral excitation energy,
and $h_{n,i}^{N}\equiv
\bra{\Psi_{0}^{N}}\sum_{\sigma}\op{n}_{i\sigma}\ket{\Psi_{n}^{N}}$
the corresponding real-valued oscillator strength.
\begin{figure*}
   \centering
   \includegraphics[height=6.5cm]{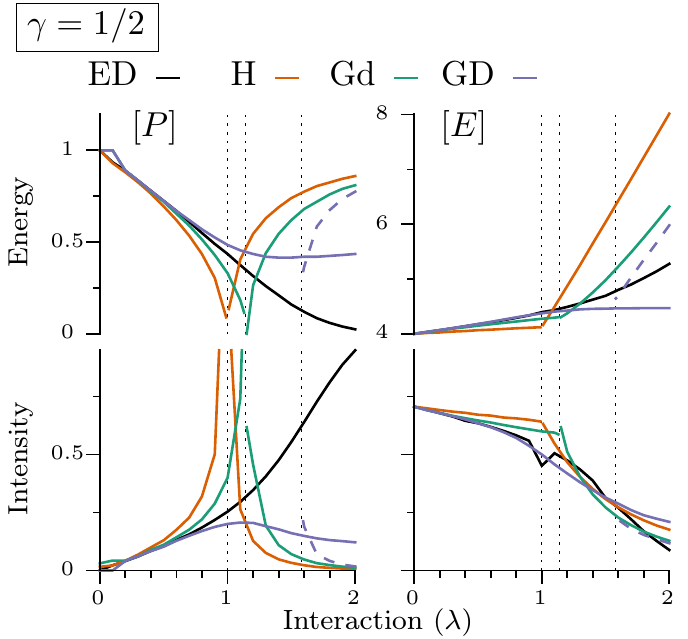}\hskip 0cm
   \includegraphics[height=6.5cm]{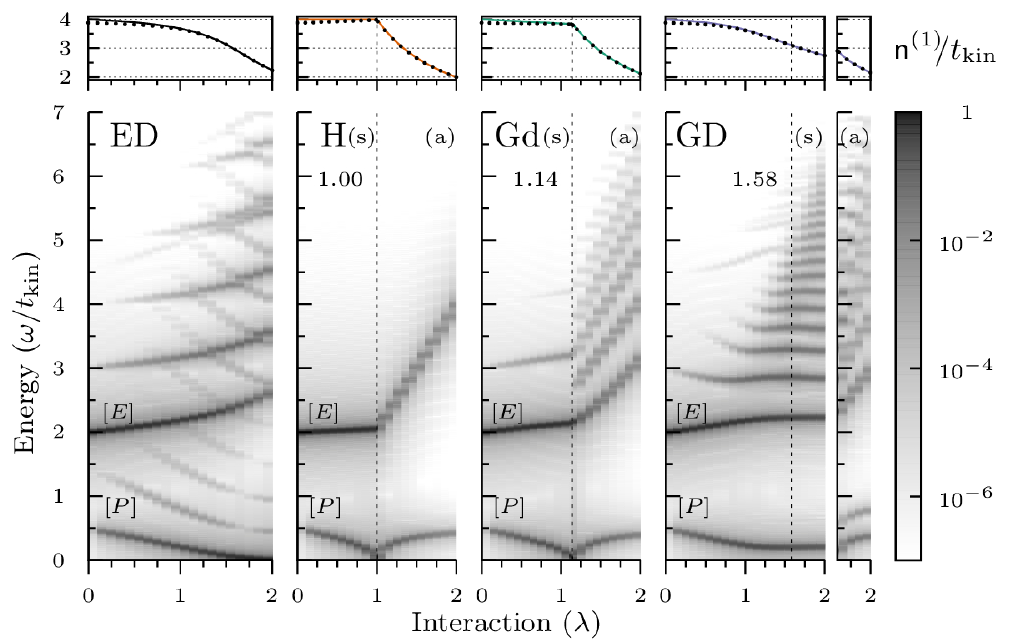}\hskip 0cm
   \includegraphics[height=6.5cm]{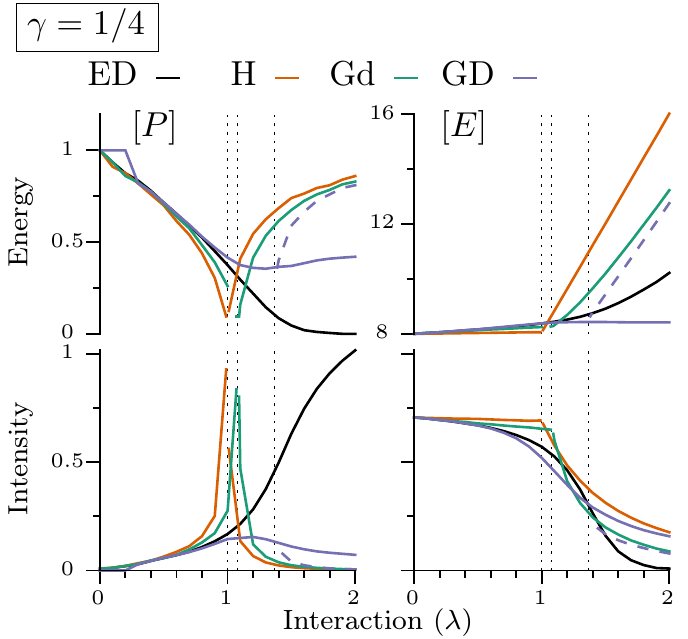}\hskip 0cm
   \includegraphics[height=6.5cm]{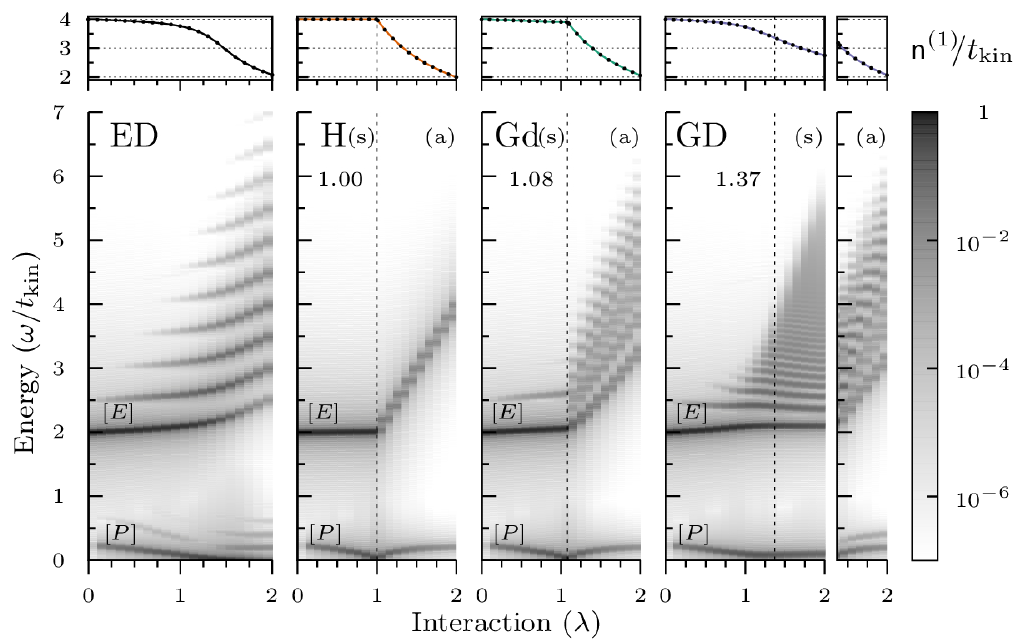}\hskip 0cm
   \caption{\label{Fig:DensityResponseFunction}
   The exact (ED) and approximate (H, Gd, GD) retarded density-density response
   function as a function of the interaction $\lambda$ and frequency $\omega$.
   The top and bottom panels correspond to the adiabatic ratios $\gamma=1/2$ and
   $\gamma=1/4$, respectively. The contour plots on the right show
   $\abs{\tr\M{\chi}^{R}(\omega)}/T$ for the stable symmetric (s) or asymmetric
   (a) solutions. The small top panels show the first $n^{(1)}$
   moment (points) and the corresponding expectation value (lines)
   to illustrate fulfillment of Eq.~\eqref{Eq:DensityResponseSumRule}. The left
   panels show the intensities and positions of the two peaks of
   $\abs{\tr\M{\chi}^{R}(\omega)}/T$ labeled with $[P]$ and $[E]$ in the contour
   plots. The stable solutions of H and Gd are denoted with a solid line
   irrespective of their symmetry. The stable symmetric and asymmetric solutions
   of GD are denoted with a solid and dashed line, respectively. (color online)}
\end{figure*}
This form is used to analyze the exact (ED) density response function
which is shown together with the approximate (H, Gd, GD) response functions in
Fig.~\ref{Fig:DensityResponseFunction}. The main panels contain contour plots
which illustrate the overall structure of the spectra. The exact results show
that the non-interacting response function
\begin{align*}
   \chi_{0;ij}^{R}(\omega)
   &\equiv\frac{(-1)^{i-j}/2}{\omega-2t_{\mathrm{kin}}+\imath\eta}
   -\frac{(-1)^{i-j}/2}{\omega+2t_{\mathrm{kin}}+\imath\eta}
   \, ,
\end{align*}
which consists of a single peak for positive frequencies, develops as a function
of the interaction to a function comprising multiple excitation energies. In the
case of weak $\lambda<1$ interactions, these excitations can be reasonably well
identified as phonon sidebands i.e.~as multi-phonon excitations either from the
non-interacting electronic ground or first, singly-excited, excited state.
The sidebands corresponding to either electronic state are separated roughly by
two bare phonon frequencies for weak interactions. Moreover, there is an
extremely faint peak located at $\omega/t_{\mathrm{kin}}\sim 4.5$ for
$\gamma=1/2$ coinciding energetically with the non-interacting doubly-excited 
electronic state plus a single phonon. As the interaction is increased, the
structure associated initially with the singly-excited electronic state labeled
with $[E]$ in the figures moves as a whole to higher energies. At the same time,
the peaks related initially to the  electronic ground state approach a bare
phonon frequency separated distribution with the lowest excitation labeled with
$[P]$ in the figures approaching zero energy and gaining intensity. These
spectral features can be understood from the adiabatic potential energy surfaces
of Fig.~\ref{Fig:PotentialEnergySurfaces} as discussed in
Sec.~\ref{Sec:PhononPropagator}. In the following, we instead focus on the
dominant low energy peak $[P]$ with the aim to identify it as a signature of a 
bipolaronic system. We start by writing the exact time-dependent density as
\begin{align*}
   n_{i}(t)
   &=\int_{-\infty}^{\infty}\!du\;\big(
   2P_{ii}(u;t)+P_{12}(u;t)\big)
   \, ,
\end{align*}
where $P_{ij}(u;t)$ defined by
\begin{align*}
	P_{ii}(u;t)
   &\equiv\lvert\langle i\uparrow,i\downarrow;u\vert
   \tilde{\Psi}_{0}^{N=2}(t)\rangle\rvert^{2}
   \, ,\\
	P_{12}(u;t)
   &\equiv\sum_{\substack{\sigma\sigma'\\\sigma\neq\sigma'}}
   \lvert\langle 1\sigma,2\sigma';u\vert
   \tilde{\Psi}_{0}^{N=2}(t)\rangle\rvert^{2}
   \, ,
\end{align*}
is the time-dependent joint probability to find the electrons at sites $i$
and $j$ and nuclei at the relative coordinate $u$ at time $t$.
Here we use the
notation $\lvert\tilde{\Psi}_{0}^{N=2}(t)\rangle
\equiv\exp(-\imath\hat{H}^{M} t)\lvert\tilde{\Psi}_{0}^{N=2}\rangle$
with $\lvert\tilde{\Psi}_{0}^{N=2}\rangle$ defined in
Eq.~\eqref{Eq:NewExactInitialState} and $\lvert i\sigma,j\sigma';u\rangle
\equiv\hat{c}^{\dagger}_{i\sigma}\hat{c}^{\dagger}_{j\sigma'}
\lvert 0\rangle_{e}\lvert u\rangle$ where $\lvert 0\rangle_{e}$ is the
electronic vacuum and $\lvert u\rangle$ is an eigenstate of $\hat{u}$.
The exact density response function can be then according to
Eq.~\eqref{Eq:RetardedDensityResponseFunction} written as
\begin{align}
\label{Eq:RetardedDensityResponseFunctionDecomposition}
   \chi^{R}_{ij}(t)
   &=2\int_{-\infty}^{\infty}\!du\;\varrho_{ii,j}^{R}(u;t)
   \, ,
\end{align}
where
\begin{align*}
   \varrho_{ik,j}^{R}(u;t)
   &\equiv\frac{\partial P_{ik}(u;t)}{\partial v_{j}}\bigg|_{v=0}
   \, ,
\end{align*}
is a response function describing how the ground state joint probability
$P_{ij}(u;0)$ changes as a function of time due to a weak perturbation. The
response function $\varrho_{12,j}(u;t)$ does not contribute to the density
response function since it is an odd function under the interchange
$u\leftrightarrow -u$ as follows from the full inversion symmetry of the model.
In Ref.~\onlinecite{saekkinen-2014a}, we use the ground state joint
probabilities as ingredients of a working definition of a dominantly bipolaronic 
ground state. The probabilities $P_{ij}(u;0)$ shown in the top panels of
Fig.~\ref{Fig:TimeDependentJointProbability} illustrate the fact that as the 
interaction increases one is most likely to find the system in a state in which
both electrons occupy the same site with an accompanying nuclear displacement.
In particular, at $\lambda=2.0$ for $\gamma=1/2$ and $\lambda=1.7$ for
$\gamma=1/4$, the ground state of the system has according to
Ref.~\onlinecite{saekkinen-2014a} crossed over to a dominantly bipolaronic
state. Next, we illustrate how these distributions behave in the linear response
regime by showing the time-average
\begin{align*}
   \langle \varrho_{ik,j}\rangle(u)
   \equiv \frac{1}{T}\int_{0}^{T}\!dt\; \abs{\varrho_{ik,j}(u;t)}
   \, ,
\end{align*}
as a function of the interaction and displacement in the left contour plots of
Fig.~\ref{Fig:TimeDependentJointProbability}. The final time $T$ is chosen here
so that $T/t_{\mathrm{kin}}^{-1}\approx470$ and
$T/t_{\mathrm{kin}}^{-1}\approx 9360$ for $\gamma=1/2$ and $\gamma=1/4$,
respectively. The results indicate that i) $\varrho_{11,1}(u,t)$ and
$\varrho_{22,1}(u,t)$ are on average larger than $\varrho_{12,1}(u,t)$, and that
for a sufficiently strong interactions the latter become suppressed while the
former gain magnitude. The maxima $\max_{u,t\in[0,T]}\abs{\varrho_{ik,j}(u;t)}$
shown in the insets underneath the averages further support these statements. 
Moreover, we observe that when the ground state distributions become spatially 
polarized as the interaction is increased, also the response functions follow
the same trend. Thus we find that ii) the spatial shapes of the initial
distributions remain qualitatively invariant in the linear response regime as a 
function of time. In order to illustrate the temporal behavior of the dominant 
response functions $\varrho_{ii,1}(u;t)$, we show them in the right panels of
Fig.~\ref{Fig:TimeDependentJointProbability} for the initially 
bipolaronic systems at $\lambda=2.0$ and $\lambda=1.7$ for $\gamma=1/2$ and
$\gamma=1/4$, respectively. Firstly, these results agree with the conclusion ii)
on the spatial structure, and secondly, they show that iii) probability density
is redistributed between $\varrho_{11,1}(u;t)$ and $\varrho_{22,1}(u;t)$ mainly
on a time-scale given by the energy scale of $[P]$, while the energy scale given
by $[E]$ is seen as superimposed small amplitude oscillations. The points i),
ii) and ii) combined allow us to conclude that, in agreement with the working
definition of Ref.~\onlinecite{saekkinen-2014a}, the system is in a dominantly 
bipolaronic state at each instant of time. Moreover, we understand the
oscillation of the probability density between $\varrho_{11,1}(u;t)$ and
$\varrho_{22,1}(u;t)$ to represent the motion of a bipolaron appearing according
to iii) on a time scale set by $[P]$. Finally, this is seen in the density 
response function according to
Eq.~\eqref{Eq:RetardedDensityResponseFunctionDecomposition}
as the emergence of the dominant low energy excitation $[P]$.

As we have described the exact density response function, we are prepared to 
investigate how the many-body approximations describe it in order to understand 
their limitations. Let us begin with the mean-field, Hartree approximation in 
which the density response function can be obtained analytically as the solution 
to the linearized Hartree equations as shown in
App.~\ref{App:HartreeDensityResponse}. This response function is given by
\begin{align*}
   \chi_{\mathrm{H}_{n};ij}^{R}(\omega)
   &\equiv\sum_{k\in\{\pm\}}
   \bigg(\frac{(-1)^{i-j}\chi_{k}^{n}/2}{\omega-\omega_{k}^{n}+\imath\eta}
   -\frac{(-1)^{i-j}\chi_{k}^{n}/2}{\omega+\omega_{k}^{n}+\imath\eta}\bigg)
   \, ,
\end{align*}
where
\begin{align*}
   \chi_{\pm}^{n}
   &\equiv
   \frac{2t_{\mathrm{kin}}\big(\omega_{0}^{2}-\omega_{\pm}^{n}{}^{2}\big)^{2}}
   {\omega_{\pm}^{n}\Big[\big(\omega_{0}^{2}-\omega_{\pm}^{n}{}^{2}\big)^{2}
   +4\lambda \omega_{0}^{2}t_{\mathrm{kin}}^{2}\Big]}
   \, ,
\end{align*}
are the oscillator strengths and
\begin{align*}
   \omega_{\pm}^{s}
   &\equiv\sqrt{\frac{\omega_{0}^{2}+4t_{\mathrm{kin}}^{2}}{2}
   \Bigg(1\pm\sqrt{1
   +\frac{16\omega_{0}^{2}t_{\mathrm{kin}}^{2}(\lambda-1)}
   {\big(\omega_{0}^{2}+4t_{\mathrm{kin}}^{2}\big)^{2}}}
   \Bigg)}
   \, ,\\
   \omega_{\pm}^{a}
   &\equiv\sqrt{\frac{\omega_{0}^{2}+4t_{\mathrm{kin}}^{2}\lambda^{2}}{2}
   \Bigg(1\pm\sqrt{1
   -\frac{16\omega_{0}^{2}t_{\mathrm{kin}}^{2}(\lambda^{2}-1)}
   {\big(\omega_{0}^{2}+4t_{\mathrm{kin}}^{2}\lambda^{2}\big)^{2}}}
   \Bigg)}
   \, ,
\end{align*}
are the frequencies for the symmetric ($s$) $\lambda<1$ and asymmetric ($a$)
$\lambda>1$ solutions, respectively. The Hartree response functions shown in the 
subpanels H of Fig.~\ref{Fig:DensityResponseFunction} thus consists of two 
contributions: the high $[E]$ ($\chi_{+}^{n}$, $\omega_{+}^{n}$) and low $[P]$
($\chi_{-}^{n}$,$\omega_{-}^{n}$) energy peaks related, for weak interactions, 
to the non-interacting electronic excited and ground states plus zero and one 
phonon, respectively. Firstly, we observe that as the interaction is increased, 
the initial distribution given by $[E]$ remains nearly invariant up to the
critical interaction $\lambda=1$ beyond which its energy increases linearly as a 
function of the interaction. Secondly, the low energy peak $[P]$, which has no 
intensity in the non-interacting $\lambda=0$ case, gains intensity and
approaches the zero energy as the interaction is increased. Moreover, when it 
reaches the zero energy at the critical interaction $\lambda=1$, its intensity 
given by the oscillator strength $\chi_{-}^{n}$ diverges. As explained in
Sec.̃\ref{Sec:Stability}, by increasing the interaction beyond this point, we 
make the symmetric equilibrium solution unstable, and therefore we change the 
initial state to one of the asymmetric solutions. This leads again to a
well-defined first order response with the intensity of the low energy peak 
becoming finite and its frequency approaching the bare phonon frequency as the  
interaction is increased. We understand these results in terms of the adiabatic
potential energy surfaces so that the mean-field approximation captures the 
lowest adiabatic potential energy surface of
Fig.~\ref{Fig:PotentialEnergySurfaces} becoming more shallow which leads to the 
lowest excitation approaching the zero energy. This agrees with the observation 
that, as the Hessian matrices of Sec.~\ref{Sec:Stability} indicate, there is a 
direction in the energy landscape in the neighborhood of the symmetric 
equilibrium solution such that along it, as $\lambda\rightarrow 1$, the 
approximately harmonic energy surface becomes more shallow. Moreover, at
$\lambda=1$, this harmonic surface becomes completely flat, and as the
linearized equations describe only this local neighborhood, it appears as if 
exciting the system costs no energy which manifests itself as the divergence of 
the zero-frequency component of the  response function. At this point, the 
lowest adiabatic potential energy surface forms the double-well structure which 
has for $\lambda=1$ also a locally flat energy landscape at $u=0$ where its 
second derivative vanishes. Lastly, one can show that the Hartree ground state energy function is equivalent to the lowest adiabatic potential energy surface
$E_{0}(u)$ by enforcing in Eq.~\eqref{Eq:HartreeTotalEnergy} that $2gn=u$ and 
that $\Gamma_{1}$ satisfies the equilibrium Hartree equations derived in
Ref.~\onlinecite{saekkinen-2014a}. This suggests that the Hartree approximation 
captures the formation of the double-well potential but needs to fall into one 
of the two minima in order to minimize the energy. By doing so, it sees again
a nearly harmonic surface, which appears in
Fig.~\ref{Fig:DensityResponseFunction} as the lowest excitation becoming finite 
and approaching the bare phonon frequency.

Let us then discuss the density response functions obtained for the stable equilibrium solutions of the partially (Gd) and fully (GD) self-consistent
Born approximations shown in Fig.~\ref{Fig:DensityResponseFunction}.
The partially self-consistent results shown in the subpanel Gd of
Fig.~\ref{Fig:DensityResponseFunction} indicate that the main qualitative
difference to the Hartree approximation is that there is a sideband structure 
related to the excitation $[E]$. The sidebands are separated roughly by two bare 
phonon frequencies in agreement with the exact results for weak interactions, 
but do not move to higher energies as a function of the interaction as clearly 
as the exact spectra does. Moreover, when $\lambda$ exceeds $\lambda_{C}$, the 
ground state becomes asymmetric and new symmetry-forbidden excitations emerge
in-between the original sidebands. In the low energy scale, we instead do not 
observe new qualitative differences to the mean-field solution, in particular we 
still find that at $\lambda_{C}$, the low energy peak $[P]$ reaches the zero 
energy with its intesity diverging. The symmetric solution of the fully
self-consistent Born approximation does, however, show a qualitative difference 
as shown in the subpanel GD (s) of Fig.~\ref{Fig:DensityResponseFunction}. In 
the low energy scale, we observe that as the interaction is increased, the low 
energy peak $[P]$ moves initially towards the zero energy but, in contrast to 
the mean-field and partially self-consistent results, does not reach it for the 
parameters considered in this work. This is in an agreement with the exact 
solution in which, however, the lowest excitation becomes increasingly close to 
the ground state, while in the fully self-consistent approximation, we observe 
that it approaches a finite non-zero value. In the high energy scale, we on the
other hand observe that as the interaction is increased, the peaks above $[E]$ 
become non-uniformly spaced and too dense in comparison to the exact solution. 
These shortcomings, as well as the fact that the spectra do not move appreciably 
to higher energies as a function of the interaction, are similar to what we 
observed for the equilibrium electron propagators in
Sec.~\ref{Sec:ElectronPropagator}. Finally, the asymmetric solutions shown in
the subpanels GD (a) are similar to the partially self-consistent solutions for
$\lambda>\lambda_{C}$ except for the additional excitation at
$\omega/\omega_{0}\sim 2$. The low energy sidebands seen in the exact solution 
are then likely merely symmetry-forbidden in the symmetric solution of the
fully self-consistent approximation. As the last remark on the overall 
structure, as shown in the top insets of Fig.~\ref{Fig:DensityResponseFunction}, 
the development of these spectra as a function of the interaction is consistent 
with the f-sum rule
\begin{align}
\label{Eq:DensityResponseSumRule}
   \mathsf{n}^{(1)}
   &\equiv-\int\limits_{-\infty}^{\infty}\!\frac{d\omega}{\pi\imath}\;
   \omega\tr\M{\chi}^{R}(\omega)
   \notag\\
   &=-2\big(E_{e}-E_{e}^{\mathrm{loc}}\big)
   -2\big(E_{ep}-E_{ep}^{\mathrm{loc}}\big)
   \, ,
\end{align}
which relates the first moment of the density-density response function to the 
electron $(E_{e},E_{e}^{\mathrm{loc}})$ and electron-phonon interaction energies 
$(E_{ep},E_{ep}^{\mathrm{loc}})$, where superscript loc refers to the
site-diagonal part of the corresponding energy. The fact that also the many-body 
approximations satisfy this sum rule is proven for the purely electronic case
in~\cite{leeuwen-2004,saekkinen-2012}.

Next, we focus on the two most significant features of the density response 
function for the weak and intermediate interactions. These are the intensity and 
position of $[P]$ and $[E]$ shown in the left panels of
Fig.~\ref{Fig:DensityResponseFunction}. In the exact case, as the interaction is 
increased, $[E]$ loses magnitude and moves towards higher energies while $[P]$ 
gains intensity and approaches the zero energy. The former is dominant up to 
borderline strong $\lambda\sim 1.5$ interactions although the latter is
appreciable already for $\lambda\sim 1$, and its impact to the response 
properties is emphasized by its different energy scale. In the Hartree 
approximation, we observe that, as a function of the interaction, $[E]$ is 
nearly invariant for $\lambda<1$ and that $[P]$ behaves in a divergent manner as 
desribed above. The mean-field approximation therefore agrees with the exact 
solution only for very weak interactions $\lambda\ll 1$. The partially 
self-consistent Born results are qualitatively similar to the mean-field  
results. On a quantitative level, it reproduces the exact intensity and
position of $[E]$ better but deviates considerably for intermediate
$\lambda\sim 1$ interactions. This together with the observed divergence of
$[P]$ implies that it can be said to agree well with exact resutls only for weak
$\lambda<1$ interactions. In both approximations, we find that the intensities
of $[E]$ and $[P]$ decrease rapidly as a function of the interaction once
$\lambda$ exceeds $\lambda_{C}$. The density response to a weak parturbation is 
thus suppressed for a sufficiently strong interaction which is consistent with 
the localized nature of the asymmetric equilibrium solutions discussed in
Ref.~\onlinecite{saekkinen-2014a}. Moreover, we find that also the asymmetric 
solution of the fully self-consistent Born approximation behaves qualitatively 
in this manner for the interactions it has been found. Lastly, the symmetric 
solution of the fully self-consistent Born approximation reproduces the exact 
positions and intensities of $[E]$ and $[P]$ well up to intermediate 
interactions $\lambda\sim 1$ with the intensity of $[E]$ being good also for 
stronger interactions. In this approximation, the low energy excitation $[P]$
does not reach the zero energy nor does it diverge as a function of the 
interaction, but its position and intensity do not still agree even
qualitatively with the exact solution for strong $\lambda>1$ interactions. 

To summarize, we have found similarly as in Sec.~\ref{Sec:ElectronPropagator}
that the Hartree, and partially and fully self-consistent Born approximations
are in a good agreement with the exact results for very weak $\lambda\ll 1$,
and weak $\lambda<1$ and up to intermediate $\lambda\sim 1$ interactions. In 
particular, we have shown that the exact density response function has, for 
sufficiently strong $\lambda>1$ interactions, a dominant low energy excitation 
which none of the approximation describe qualitatively correctly. Moreover,
we have related this excitation to the response of a bipolaron to a weak  perturbation by analyzing it using the time-dependent joint probabilities.
Instead of describing the low energy excitation, the Hartree and partially
self-consistent Born approximations give rise to a divergence of the response 
function at the critical point $\lambda_{C}$. This has been explained by
relating it to the formation of the double-well structure in the lowest
adiabatic potential energy surface which we have also associated with the 
bipolaronic crossover in Ref.~\onlinecite{saekkinen-2014a}. Finally, we have
shown that the divergence can be prevented by dressing the phonon propagator
self-consistently at the level of the fully self-consistent Born approximation.


\subsubsection{Phonon Propagator Revisited}
\begin{figure}
   \includegraphics[scale=0.7]{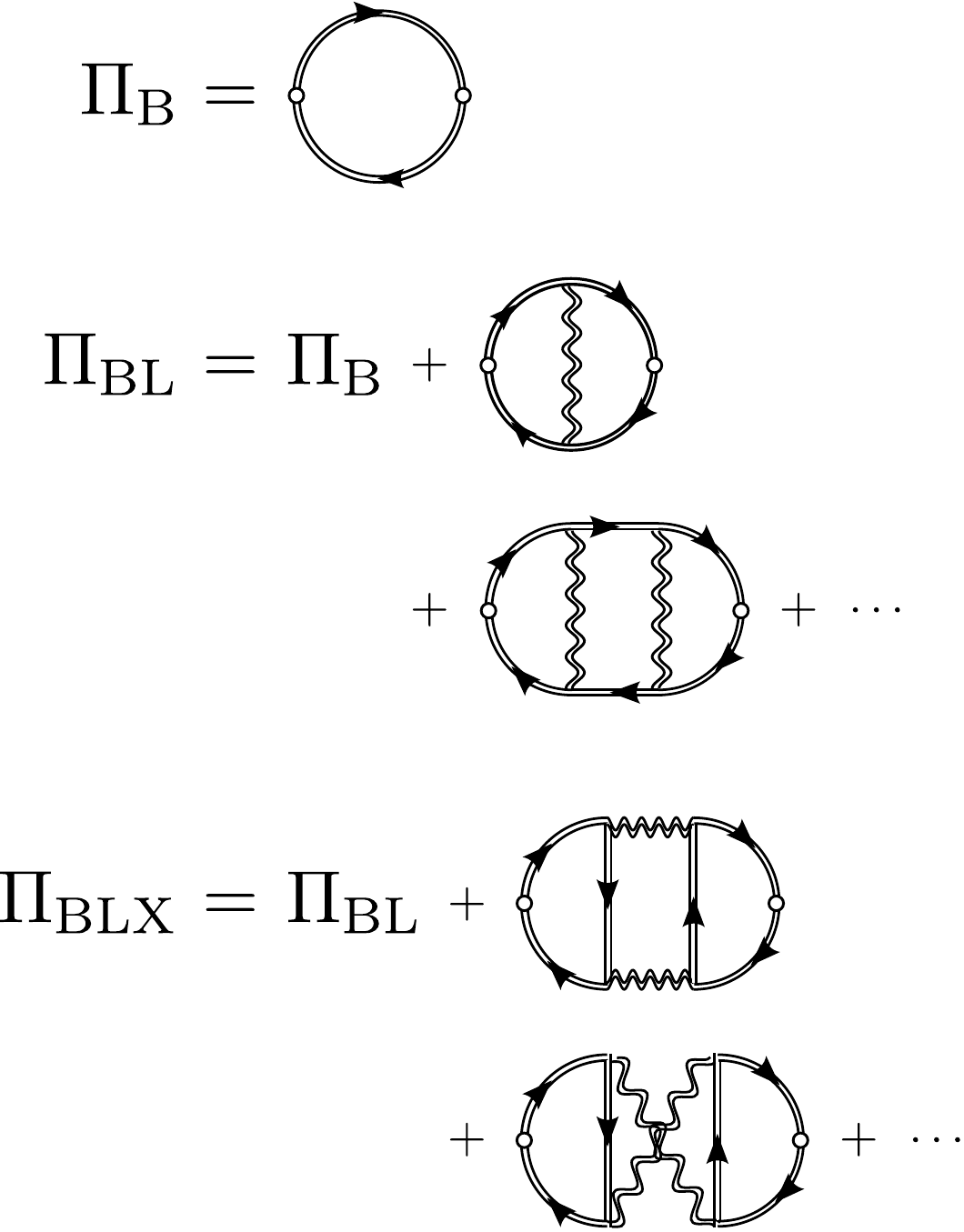}
   \caption{\label{Fig:phonon-selfenergy}
   The phonon self-energies, or their functional forms, corresponding to
   the phonon propagators obtained by time-propagation. The Hartree (H), and
   partially (Gd) and fully (GD) self-consistent Born approximations
   relate to the bubble $(B)$, and bubble-ladders (BL) and
   bubble-ladders-exchange (BLX) self-energy functionals, respectively. A line 
   with an arrow indicates a dressed electron propagator, while single and
   two-fold wiggly lines represent bare and dressed phonon propagators, 
   respectively. An open circle represents a connection for a phonon
   propagator.}
\end{figure}

The density response function which we calculated and discussed above describes
electron density fluctuations which in turn couple to nuclear density
fluctuations described by the phonon propagator. This is formally shown by
the Dyson equation of Eq.~\eqref{Eq:DysonEquationPhononPropagator} which
can be written as
\begin{align}
\label{Eq:DysonEquationReducible}
   \M{D}(z;z')
   &=\M{d}(z;z')
   \notag\\
   &+\int\limits_{C}\!d\bar{z}d\bar{z}'\;
   \M{d}(z;\bar{z})\M{\Pi}_{r}(\bar{z};\bar{z}')\M{d}(\bar{z}';z')
   \, ,\\
   \Pi_{r,PQ}(z;z')
   &\equiv \sum_{ijkl}M_{ji}^{P}(z)\chi_{ij,kl}(z;z')M_{lk}^{Q}(z')
   \, ,
\end{align}
where the reducible self-energy $\Pi_{r}$ is determined by the generalized
response function of Eq.~\eqref{Eq:GeneralizedResponseFunction}. This means that
a density response function obtained by time-propagation can be also used to
obtain a new phonon propagator. Here we use this relation to identify some
phonon self-energies and discuss whether or not they lead to better nuclear
properties than the fully self-consistent Born approximation (GD). 
In order to do this, due to computational reasons instead of using
the equilibrium frequency-domain version of Eq.~\ref{Eq:DysonEquationReducible},
we perturb the system with the instantaneous force
\begin{align*}
   \op{F}_{P}(t)
   &=\delta(t)F_{P}
   \, ,
\end{align*}
where $F_{P}$ is the magnitude of the perturbation. We then record the resulting
phonon field expectation value which in the linear response regime satisfies
\begin{align*}
   \delta \phi_{P}(t)
   &\equiv \phi_{P}(t)-\phi_{P}^{(0)}(t)
   \notag\\
   &=\sum_{Q}D^{R}_{PQ}(t)F_{Q}
   +\mathcal{O}(F^{2}) \, ,
\end{align*}
where $\phi_{P}^{(0)}(t)$ is the expectation value of the unperturbed system,
and $D^{R}_{PQ}(t)$ is the retarded phonon propagator. In the Kadanoff-Baym
equations this perturbation amounts to choosing
\begin{align*}
   \phi_{P}(0)
   &=\phi_{P}^{M}-\imath\sum_{Q}\alpha_{PQ}F_{Q}
   \, ,
\end{align*}
where $\phi_{P}^{M}$ is the equilibrium expectation value, as the new initial
condition and subsequently solving the equations of motion in the absence of
this perturbation. The phonon propagator is then given by
\begin{align*}
   D_{PQ}^{R}(t)
   &\equiv\frac{\partial\phi_{P}(t)}{\partial F_{Q}}\bigg|_{F=0}
   \, .
\end{align*}
which we in practice evaluate by using the difference quotient
$(\phi_{P}(t)-\phi_{P}^{(0)}(t))/F_{Q}$ with sufficiently small $F_{Q}$ and
$F_{R}=0$ for $R\neq Q$. It can be shown that this propagator satisfies
Eq.~\eqref{Eq:DysonEquationReducible} and its irreducible version of Eq.~\eqref{Eq:DysonEquationPhononPropagator} with the irreducible self-energy
\begin{align*}
   \Pi_{PQ}(z;z')
   &= \sum_{ijkl}M_{ji}^{P}(z)P_{ij,kl}(z;z')M_{lk}^{Q}(z')
   \, ,
\end{align*}
where $P$ is the irreducible  polarizability of Eq.~\eqref{Eq:polarizability}.
The phonon propagators obtained by time-propagation are thus related to
irreducible self-energy functionals whose lowest-order diagrammatic expansions
are shown in Fig.~\ref{Fig:phonon-selfenergy}. The phonon propagator obtained
in this manner in the Hartree (H), and partially (Gd) and fully (GD)
self-consistent Born approximations are respectively given by
\begin{align*}
   \Pi_{\mathrm{td-H}}(z;z')
   &=\Pi_{\mathrm{B}}[G_{\mathrm{H}},d](z;z')
   \, ,\\
   \Pi_{\mathrm{td-Gd}}(z;z')
   &=\Pi_{\mathrm{BL}}[G_{\mathrm{Gd}},d](z;z')
   \, ,\\
   \Pi_{\mathrm{td-GD}}(z;z')
   &=\Pi_{\mathrm{BLX}}[G_{\mathrm{GD}},D_{\mathrm{GD}}](z;z')
   \, ,   
\end{align*}
where we have introduced the prefix 'td-' referring ot time-dependent to
distinguish these self-energies from their original counterparts. This shows
explicitly that these approximations are not self-consistent i.e.~the propagator
satisfying the Dyson equation is not the same as the ones in the self-energy
diagrams.

\begin{figure}
   \centering
   \includegraphics[height=3cm]{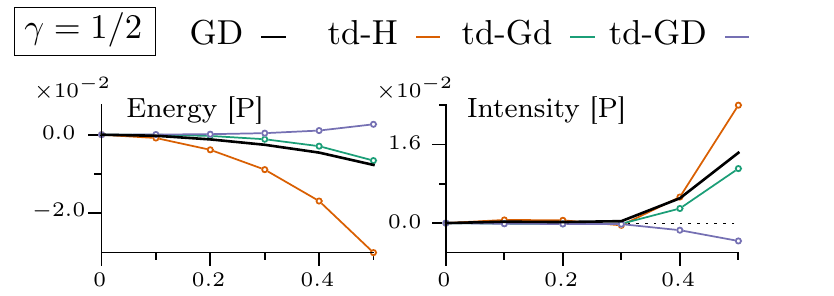}\vskip 0cm
   \includegraphics[height=6.5cm]{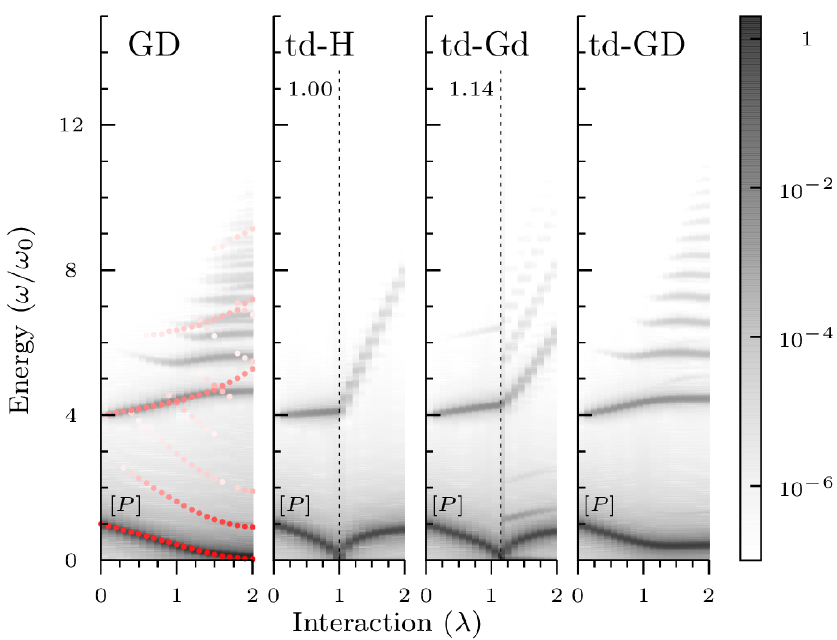}\vskip 0cm
   \includegraphics[height=3cm]{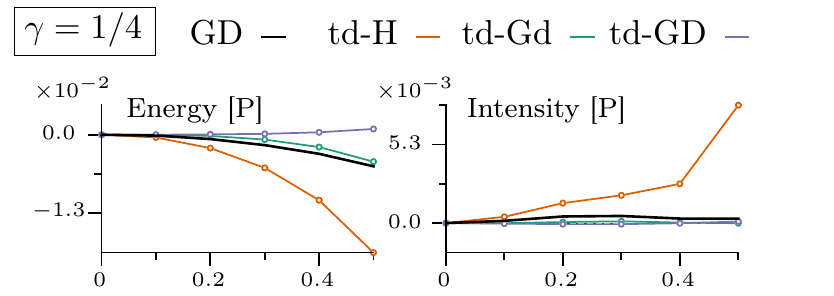}\vskip 0cm
   \includegraphics[height=6.5cm]{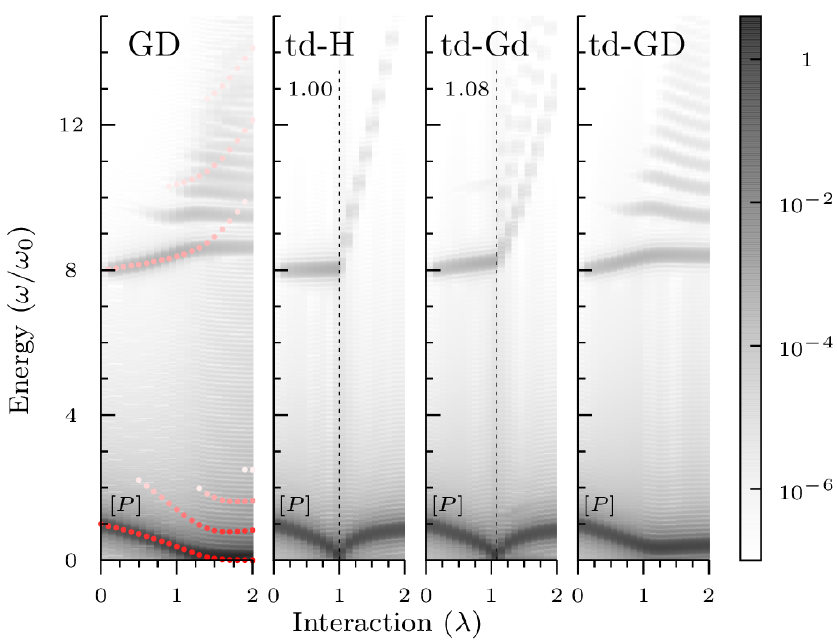}
   \caption{\label{Fig:DisplacementResponseFunction}
   The phonon propagators as a function of the interaction $\lambda$ and
   frequency $\omega$. The top and bottom figures correspond to the adiabatic
   ratios $\gamma=1/2$ and $\gamma=1/4$, respectively. The contour plots show
   $\abs{D_{11}^{R}(\omega)}/T$ for the exact (ED), denoted with red
   dots, and approximate (GD, td-H, td-Gd, td-GD) solutions. The top panels show
   the difference between the approximate and exact intensity and position of
   the lowest energy peak of $\abs{D_{11}^{R}(\omega)}/T$ labeled with $[P]$ in
   the contour plots. (color online)}
\end{figure}

The results can be anticipated by noting that the non-interacting phonon
propagator is a function peaked at the bare phonon frequencies. The reducible
frequency-domain Dyson equation then suggests that the phonon propagator
satisfying it has a similar frequency content as the density response function
with weight redistributed around the bare phonon frequencies. This already gives
a picture how good are the phonon propagators obtained from the density response
functions of Sec.~\ref{Sec:DensityDensityResponseFunction}. However, let us try
to make this picture more quantitative. Figure~\ref{Fig:DisplacementResponseFunction} shows
the Fourier transforms of the retarded phonon propagators obtained by
time-propagation for the many-body approximations (td-H, td-Gd, td-GD). The
exact (ED) and fully self-consistent Born (GD) equilibrium propagators are also
shown for reference. We have discussed the reference results and their physical
content in Sec.~\ref{Sec:PhononPropagator} so here we focus directly to
comparing the different approximations. The contour plots show that none of the
new approximations improve the qualitative description of the low energy peak
$[P]$ which dominates the spectra. Moreover, only td-Gd with a symmetry-broken
ground state for $\lambda>\lambda_{C}$ produces a sideband structure for this
excitation. We also observe that td-GD has a slightly larger sideband separation
for the electronic excitation when compared to the fully self-consistent Born
(GD) spectra. Here we remark that the numerical results for td-H agree with the
analytical results and discussion on the phonon vacuum instability presented
in Ref.~\onlinecite{saekkinen-2014a}. Lastly, we show the position and intensity
of $[P]$ relative to its exact position and intensity in the top panels of
Fig.~\ref{Fig:DisplacementResponseFunction} for weak interactions $\lambda<0.5$.
The results highlight, as expected in a perturbative regime, that td-H deviates
the most and td-GD the least from the exact result. Furthermore td-Gd and GD
give similar results in this regime with the former being slightly better as it
includes all the diagrams up to fourth order in the electron-phonon interaction.

Overall due to the poor description of the lowest excitation, the td-H, td-Gd,
and td-GD approximations are only valid in the regime of weak interactions
$\lambda<1$ in which td-Gd and td-GD improve on GD. The qualitative behavior
for larger interactions $\lambda\sim 1$ shows that although one can obtain
sophisticated many-body self-energies by means of time-propagation, they do not
necessarily improve the description of the physics. In particular, we find that
infinite summation schemes for self-energy diagrams, which include vertex
corrections, lead to deterioration of the spectral properties of
the equilibrium propagator evaluated with a single dressed polarization bubble
for intermediate to high interactions.

\section{Conclusions and Outlook}
\label{Sec:Conclusions}
We have introduced a method based on time-dependent many-body perturbation
theory aimed at studying interacting electrons and phonons. The many-body
approximations used here are the Hartree (H), and the partially (Gd) and fully
(GD) self-consistent Born approximations. The method has been applied to
investigate both the non-neutral and neutral excitation spectra of a two-site,
two-electron Holstein model. We have presented results for the frequency-domain
ground-state electron and phonon propagators, as well as for the density-density
and displacement-displacement linear response functions. The results have been 
compared with numerically exact results obtained by exact diagonalization (ED)
in order to assess their quality and relate a physical picture to the
behavior of the many-body approximations.

In Ref.~\onlinecite{saekkinen-2014a}, we found that the approximations studied
here support multiple, concurrently co-existing solutions some of which exhibit
a broken reflection symmetry. The asymmetric solutions were found once the
electron-phonon interaction $\lambda$ reached a critical value $\lambda_{C}$,
and were understood to mimic the bipolaronic crossover of the exact solution.
The asymmetric solutions were furthermore found to be the lowest energy
solutions for a large range of parameters. The total energies, and natural
occupation numbers, also suggested that the symmetric solution of the fully
self-consistent Born approximation describes partially the crossover to a
bipolaronic state. In the present work, we studied the frequency structure of
the ground state propagators for a restricted range of adiabatic ratios
$\gamma=1/2,1/4$ and interactions $\lambda\in[0,2]$ which allows us to complete
some of the observations made in~Ref.~\onlinecite{saekkinen-2014a}. Firstly,
the frequency-integrated observables obtained from the electron propagator are
in a better qualitative agreement with exact results than the frequency-resolved 
objects themselves. In particular, our results show that none of the
approximations give an electron propagator in which there is a rigid shift or 
redistribution of the  spectral weight comparable to the exact solution for
high $\lambda>1$ interactions. The phonon propagator does not moreover show a
clear spectral fingerprint of a double-well structure in the fully
self-consistent Born approximation for the parameters considered. As all of
these features have been identified as spectral signatures of a bipolaronic
state, our results here favor the statement that none of the approximations
describe even partially the bipolaronic crossover for the parameters considered.
This said, the results for the electron propagator can be roughly summarized by 
concluding that the Hartree, partially self-consistent Born, and fully
self-consistent Born approximations agree with the exact results up to very weak
$\lambda\ll 1$, weak $\lambda< 1$, and weak to intermediate $\lambda\sim 1$ 
interactions, respectively. The non-interacting phonon propagator used in the
Hartree and partially self-consistent Born approximations is only valid for very
weak $\lambda\ll 1$ interactions while the dressed propagator of the fully
self-consistent Born approximation agrees reasonably well with the exact results
up to borderline strong $\lambda\sim 1.5$ interactions.

The linear response functions studied in the present work are obtained by
time-propagation starting either from a symmetric or an asymmetric ground state solution. In the case $\lambda>\lambda_{C}$, these solutions co-exits and we
find that the symmetric solutions of the Hartree and partially self-consistent
Born approximations are unstable, while the asymmetric solutions are stable,
against a small asymmetric perturbation. The symmetric and asymmetric solutions
of the fully self-consistent Born approximation are, on the other hand, shown
to be stable against the same perturbation. By identifying the stable ground
state solutions, we have been able to evaluate linear response functions
corresponding to these solutions. In particular, the density-density response
function obtained in the Hartree and partially self-consistent Born
approximations is shown to have a zero-frequency component which appears and its
intensity diverges as $\lambda$ approaches $\lambda_{C}$. In the Hartree 
approximation, this is caused by the build-up of a double-well structure in its 
ground state energy surface exactly at $\lambda=1$ and by the need of the
mean-field to minimize the total energy. Our results further show that the fully
self-consistent Born approximation does not have a similar divergence.
The comparison of the exact and approximate density-density response functions
confirms that the range of validity of the many-body approximations is roughly
the same as for the case of equilibrium propagators. In particular, none of the
many-body approximations were able to describe the lowest excitation of the
exact response function for strong interactions $\lambda>1$ for which it is the 
dominant feature of the exact response function. By analyzing this excitation,
we further identified it as a signature of a bipolaronic system, and hence in 
agreement with the conclusions based on the equilibrium propagators, this
suggests that the approximations do not describe the crossover to the
bipolaronic state. Finally, we could by time-propagation obtain another phonon 
propagator associated with a highly sophisticated self-energy, although a
non-selfconsistent one. The results show that although the propagators obtained
in the partially and fully self-consistent approximations are better than the
equilibrium propagator of the fully self-consistent Born approximation for the perturbative $\lambda\ll 1$ interactions, they lead to deterioration of the 
qualitative spectral properties for higher interactions. This suggests that,
at least in this case, it is either important to maintain self-consistency,
or that instead of infinite partial summations of dressed self-energy diagrams
it is better to consider truncated approximations.

In order to go beyond the approximations studied here in many-body perturbation
theory one needs to introduce vertex corrections which are in particular needed
to improve the properties of the electron propagator. This is realizable in an 
equilibrium theory but leads to a substantial increase in the complexity of the 
time-domain method which makes the inclusion of vertex corrections challenging 
with the current computational resources. The two-time equations can be however 
cast as one-time equations via the GKBA~\cite{kohler-1996,bonitz-1996,
gartner-2006,pavlyukh-2009,pavlyukh-2011,balzer-2013,latini-2014,marini-2013,
sangalli-2015a,sangalli-2015b} which is a possible way to overcome the numerical 
challenge and reduce more advanced approximations tractable. On the other hand 
e.g.~in cavity quantum electrodynamics already our weak interactions are 
considered strong and thus the approximations used here could be a valuable
asset for investigating non-linear time-dependent phenomena. In this context our 
results provide a basis for studying the approximation in an explicitly
time-dependent situation, and for understanding properties of further approximations e.g.~the GKBA which could be used in order to address physics of 
larger or more realistic systems.


\begin{acknowledgments}
We would like to thank Daniel Karlsson, Riku Tuovinen, and Christian 
Sch\"afer for careful reading of the manuscript and useful discussions.
We acknowledge CSC – IT Center for Science Ltd for the allocation of
computational resources. RvL acknowledges the Academy of Finland for support
under grant No.~267839.
\end{acknowledgments}


\appendix

\section{Hartree, TDSCF and Ehrenfest}
\label{App:HartreeTdscfEhrenfest}
The Hartree approximation is shown here to be equivalent to the time-dependent
self-consistent field (TDSCF) approach and to the semi-classical Ehrenfest
approximation. This equivalence is shown by deriving the time-dependent
self-consistent field equations, noting that they reduce to the
Ehrenfest equations of motion, and finally by showing that their solutions
can be used to construct the phonon field expectation value, and the electron
propagator of the Hartree approximation. We begin by introducing the product ansatz
\begin{align*}
   \ket{\Psi(t)}
   &\equiv\ket{\tilde{\psi}(t)}\ket{\tilde{\chi}(t)}
   \, ,
\end{align*}
where $\ket{\tilde{\psi}(t)}$ and $\ket{\tilde{\chi}(t)}$ consist of only electronic
and phononic degrees of freedom, respectively. By substituting this ansatz to
the time-dependent Schr\"{o}dinger equation
\begin{align*}
   \imath\partial_{t}\ket{\Psi(t)}
   &=\hat{H}(t)\ket{\Psi(t)}
\end{align*}
and projecting it to the states $\ket{\tilde{\chi}(t)}$ and
$\ket{\tilde{\psi}(t)}$, we arrive at the time-dependent self-consistent field
equations~\cite{marx-book}
\begin{align*}
   \imath\partial_{t}\ket{\psi(t)}
   &=\sum_{ij}\Bigg(h_{ij}(t)
   +\sum_{P}M_{ij}^{P}(t)\phi_{P}(t)\Bigg)
   \hat{c}_{i}^{\dagger}\hat{c}_{j}\ket{\psi(t)}
   \, ,\\
   \imath\partial_{t}\ket{\chi(t)}
   &=\Bigg(\sum_{PQ}\Omega_{PQ}(t)\hat{\phi}_{P}(t)\hat{\phi}_{Q}(t)
   \notag\\
   &+\sum_{P}F_{P}(t)\hat{\phi}_{P}(t)
   +\sum_{ijP}M_{ij}^{P}(t)\gamma_{ji}(t)\hat{\phi}_{P}\Bigg)\ket{\chi(t)}
\end{align*}
where
\begin{align*}
   \phi_{P}(t)
   \equiv\bra{\chi(t)}\op{\phi}_{P}\ket{\chi(t)}
   \, ,\\
   \gamma_{ij}(t)
   \equiv\bra{\psi(t)}\op{c}_{j}^{\dagger}\op{c}_{i}\ket{\psi(t)}
   \, ,
\end{align*}
are the phonon field expectation value and the reduced density
matrix. In the derivation. we adopted the phase conventions
\begin{align*}
   \ket{\psi(t)}
   &\equiv e^{\imath\int_{t_{0}}^{t}\!dt'\;
   \big(E_{p}(t')
   -\langle\tilde{\chi}(t')\vert\imath\partial_{t'}\tilde{\chi}(t')\rangle\big)}
   \ket{\tilde{\psi}(t)}
   \, ,\\
   \ket{\chi(t)}
   &\equiv e^{\imath\int_{t_{0}}^{t}\!dt'\;
   \big(E_{e}(t')
   -\langle\tilde{\psi}(t')\vert\imath\partial_{t'}\tilde{\psi}(t')\rangle\big)}
   \ket{\tilde{\chi}(t)}
   \, ,
\end{align*}
where we defined
\begin{align*}
   E_{e}(t)
   &\equiv\sum_{ij}h_{ij}(t)
   \bra{\psi(t)}\hat{c}_{i}^{\dagger}\hat{c}_{j}\ket{\psi(t)}
   \, ,\\
   E_{p}(t)
   &\equiv\sum_{PQ}\Omega_{PQ}(t)
   \bra{\chi(t)}\hat{\phi}_{P}\hat{\phi}_{Q}\ket{\chi(t)}
   \, .
\end{align*}
as the electron and phonon energies, respectively. The semi-classical Ehrenfest
equations are then derived e.g.~by introducing a polar expansion of the nuclear
state and taking its classical limit~\cite{marx-book}. In our case
however, the Heisenberg equations of motion for the phonon field expectation
values are equal to the classical equations of motion. Taking advantage of this
property and the bilinearity of the equation for $\ket{\psi(t)}$ in terms of the
electronic operators, we arrive at the Ehrenfest equations of motion
\begin{subequations}
\label{Eq:EhrenfestEquations}
\begin{align}
   \imath\partial_{t}\psi_{ij}(t)
   &=\sum_{k}\Bigg(h_{jk}(t)
   \notag\\
   &+\sum_{P}M_{jk}^{P}(t)\phi_{P}(t)\Bigg)\psi_{ik}(t)
   \, ,\\
\label{Eq:EhrenfestEquationB}
   \imath\sum_{Q}\alpha_{PQ}\partial_{t}\phi_{Q}(t)
   &=\sum_{Q}\tilde{\Omega}_{PQ}(t)\phi_{Q}(t)
   \notag\\
   &+F_{P}(t)
   +\sum_{ij}M_{ij}^{P}(t)\gamma_{ji}(t)
   \, ,
\end{align}
\end{subequations}
such that $\ket{\psi(t)}$ can be written as a Slater determinant of the
time-dependent orbitals $\psi_{i}(t)$. In order to relate these orbitals to the
electron propagator in the Hartree approximation, we further write down the
equilibrium Hartree equations
\begin{subequations}
\label{Eq:EquilibriumHartreeEquations}
\begin{align}
   \M{h}_{\mathrm{H}}^{M}\V{\psi}_{k}^{M}
   &=\epsilon_{k}^{M}\V{\psi}_{k}^{M}
   \, ,\\
   \M{h}_{\mathrm{H}}^{M}
   &=\M{h}^{\mathrm{M}}
   +\sum_{P}\M{M}^{P}\phi_{P}^{M}
   \, ,\\
   \V{\phi}^{M}
   &=-\M{\tilde{\Omega}}^{M}{}^{-1}\bigg(\V{F}^{\mathrm{M}}
   +\sum_{ij}\V{M}_{ij}\gamma_{ji}^{M}\bigg)
   \, ,
\end{align}
\end{subequations}
which are introduced in Ref.~\onlinecite{saekkinen-2014a}, and correspond to a
set of non-linear eigenvalue equations for the eigenvalues $\epsilon_{k}^{M}$
and eigenvectors $\psi_{k}^{M}$. The electron propagator can be then written in
the Hartree approximation in terms of the time-dependent orbitals $\psi_{i}(t)$
obtained by solving Eqs.~\eqref{Eq:EhrenfestEquations} with $\phi_{P}^{M}$ and
$\psi_{i}^{M}$ as their initial conditions. That is, in the Hartree
approximation, the phonon field expectation values satisfy
Eq.~\eqref{Eq:EhrenfestEquationB}, and the electron propagator is given by
\begin{align*}
   G_{ij}^{>}(t;t')
   &=\frac{1}{\imath}\sum_{k}
   \bar{f}_{+}\big(\beta\epsilon_{k}^{M}\big)
   \psi_{kj}^{*}(t')\psi_{ki}(t)
   \, ,\\
   G_{ij}^{<}(t;t')
   &=-\frac{1}{\imath}\sum_{k}
   f_{+}\big(\beta\epsilon_{k}^{M}\big)
   \psi_{kj}^{*}(t')\psi_{ki}(t)
   \, ,
\end{align*}
where $\bar{f}_{+}\equiv 1-f_{+}$, as readily verified by using Eqs.~\eqref{Eq:EhrenfestEquations} to check that Eqs.~\eqref{Eq:EquationsOfMotion} with the
Hartree self-energy are satisfied, and also by verifying that the
Kubo-Martin-Schwinger boundary conditions~\cite{stefanucci-book} are met.


\section{Hartree Density Response Function}
\label{App:HartreeDensityResponse}
Here, we calculate the density response function of the two-site,
two-electron Holstein model in the Hartree approximation by applying the method
discussed in Sec.~\ref{Sec:ResultsMethod}. In order to do this, we first rewrite
the Hartree equations in a more convenient form by using the conserved total
energy to reduce the number of dependent variables. That is, the total energy of
Eq.~\eqref{Eq:HartreeTotalEnergy} allows us to eliminate $\Gamma_{1}$, and
subsequently by defining the vector
\begin{align*}
   \V{x}&\equiv
   \begin{pmatrix} n \\ \Gamma_{2} \\ u \\ p \end{pmatrix}
   \, ,
\end{align*}
we can rewrite the Hartree equations given in
Eqs.~\eqref{Eq:HartreeEquationsForDimer} as
\begin{align}
\label{Eq:HartreeEquationsVectorForm}
   \dot{\V{x}}
   &=\V{f}(\V{x})
   \notag\\
   &\equiv
   \begin{pmatrix}
      4t_{\mathrm{kin}}x_{2} \\
      -t_{\mathrm{kin}}x_{1}
      +2gx_{3}\Gamma_{1}(x_{1},x_{3},x_{4}) \\
      \omega_{0}x_{4} \\
      -\omega_{0}x_{3}+2gx_{1}
   \end{pmatrix}
   \, ,
\end{align}
where $8t_{\mathrm{kin}}\Gamma_{1}(x_{1},x_{3},x_{4})
\equiv\omega_{0}\big(x_{4}^{2}+x_{3}^{2}-1\big)-4gx_{1}x_{3}-2E_{0}$.
Here $E_{0}$ denotes the total energy at $t=0$ which is determined by
the initial condition $\V{x}^{0}$. The density-density response function of
Eq.~\eqref{Eq:RetardedDensityResponseFunction}, which is in this case given by
\begin{align}
\label{Eq:HartreeDensityResponseFunction}
   \chi_{ij}^{R}(t)
   &=(-1)^{i+1}\frac{\partial n(t)}{\partial v_{j}}\bigg|_{v=0}
   \, .
\end{align}
is then the $i=1$ component of the more general response function
$\partial x_{i}/\partial v_{j}|_{v=0}$ which satisfies
\begin{align*}
   \frac{d}{dt}\frac{\partial \V{x}}{\partial v_{j}}\bigg|_{v=0}
   &=\M{J}\frac{\partial \V{x}}{\partial v_{j}}\bigg|_{v=0}
\end{align*}
as seen by differentiating Eq.~\eqref{Eq:HartreeEquationsVectorForm} with
respect to $v_{j}$. The Jacobian matrix
$J_{ij}\equiv\partial_{x_{j}}f_{i}(\V{x})|_{v=0}$ is a function of the
unperturbed solution $\V{x}|_{v=0}$ which is in our case obtained by propagating
either the symmetric or asymmetric ground state solution of the equilibrium
Hartree equations given in Eq.~\eqref{Eq:SymmetricHartreeFixedPoint} and
Eq.~\eqref{Eq:AsymmetricHartreeFixedPoint}, respectively. As these solutions are
also fixed-points of Eq.~\eqref{Eq:HartreeEquationsVectorForm}, they are
constant in time, and thus the Jacobian matrices for the symmetric ($s$) and
both asymmetric ($a$) solutions given by
\begin{align*}
   \M{J}_{s}&\equiv
   \begin{pmatrix}
      0 & 4t_{\mathrm{kin}} & 0 & 0 \\
      -t_{\mathrm{kin}} & 0 & g & 0 \\
      0 & 0 & 0 & \omega_{0} \\
      2g & 0 & -\omega_{0} & 0
   \end{pmatrix}
   \, ,\\
   \M{J}_{a}&\equiv
   \begin{pmatrix}
      0 & 4t_{\mathrm{kin}} & 0 & 0 \\
      -t_{\mathrm{kin}}\lambda^{2} & 0 & g\lambda^{-1} & 0 \\
      0 & 0 & 0 & \omega_{0} \\
      2g & 0 & -\omega_{0} & 0
   \end{pmatrix}
   \, ,
\end{align*}
are time-independent. The initial conditions
$\partial \V{x}_{\eta}^{0}/\partial v_{j}|_{v=0}$ for the symmetric ($\eta=s$)
and asymmetric ($\eta=a$) cases can be deduced from
Eqs.~\eqref{Eq:NewApproximateInitialState} to be
\begin{align*}
   \frac{\partial \V{x}_{s}^{0}}{\partial v_{j}}\bigg|_{v=0}
   &=\begin{pmatrix}
      0 \\ -(\delta_{1j}-\delta_{2j})/2 \\ 0 \\ 0
   \end{pmatrix}
   \, .\\
   \frac{\partial \V{x}_{a}^{0}}{\partial v_{j}}\bigg|_{v=0}
   &=\begin{pmatrix}
      0 \\ -(\delta_{1j}-\delta_{2j})\lambda^{-1}/4 \\ 0 \\ 0
   \end{pmatrix}
   \, .
\end{align*}
respectively. The equation for the response function is linear, and hence admits
the solution
\begin{align*}
   \frac{\partial \V{x}_{\eta}(t)}{\partial v_{j}}\bigg|_{v=0}
   &=e^{\M{J}_{\eta}t}\frac{d\V{x}_{\eta}^{0}}{dv_{j}}\bigg|_{v=0}
   \, ,
\end{align*}
where we restored the explicit time-dependence. The task is then to evaluate the
matrix exponential which is done here by using the eigendecomposition of the
Jacobian matrix. The decomposition exists since the eigenvalues of the Jacobian
matrix given by
$\imath\omega_{\pm}^{\eta}$, $-\imath\omega_{\pm}^{\eta}$, where
\begin{align*}
   \omega_{\pm}^{s}
   &\equiv\sqrt{\frac{\omega_{0}^{2}+4t_{\mathrm{kin}}^{2}}{2}
   \Bigg(1\pm\sqrt{1
   +\frac{16\omega_{0}^{2}t_{\mathrm{kin}}^{2}(\lambda-1)}
   {\big(\omega_{0}^{2}+4t_{\mathrm{kin}}^{2}\big)^{2}}}
   \Bigg)}
   \, ,\\
   \omega_{\pm}^{a}
   &\equiv\sqrt{\frac{\omega_{0}^{2}+4t_{\mathrm{kin}}^{2}\lambda^{2}}{2}
   \Bigg(1\pm\sqrt{1
   -\frac{16\omega_{0}^{2}t_{\mathrm{kin}}^{2}(\lambda^{2}-1)}
   {\big(\omega_{0}^{2}+4t_{\mathrm{kin}}^{2}\lambda^{2}\big)^{2}}}
   \Bigg)}
   \, ,
\end{align*}
are non-degenerate for $\lambda\neq 1$. The diagonalizing similarity
transformation given by
\begin{align*}
   \M{X}^{\eta}
   &\equiv
   \begin{pmatrix}
      1 & 1 & 1 & 1 \\
      \imath\omega_{+}^{\eta}/4t_{\mathrm{kin}} &
      \imath\omega_{-}^{\eta}/4t_{\mathrm{kin}} &
      -\imath\omega_{+}^{\eta}/4t_{\mathrm{kin}} &
      -\imath\omega_{-}^{\eta}/4t_{\mathrm{kin}} \\
      x_{+}^{\eta} &
      x_{-}^{\eta} &
      x_{+}^{\eta} &
      x_{-}^{\eta} \\ 
      \imath\omega_{+}^{\eta}x_{+}^{\eta} &
      \imath\omega_{-}^{\eta}x_{-}^{\eta} &
      -\imath\omega_{+}^{\eta}x_{+}^{\eta} &
      -\imath\omega_{-}^{\eta}x_{-}^{\eta} 
   \end{pmatrix}
   \, ,
\end{align*}
where $x_{\pm}^{\eta}\equiv 2g/\big(1-\omega_{\pm}^{\eta}{}^{2}\big)$, then
allows us to write the response function as
\begin{align*}
   \frac{\partial \V{x}_{\eta}(t)}{\partial v_{j}}\bigg|_{v=0}
   &=\M{X}e^{\imath\M{\omega}^{\eta}t}\M{X}^{-1}
   \frac{\partial \V{x}_{\eta}^{0}}{\partial v_{j}}\bigg|_{v=0}
   \, ,
\end{align*}
where $\M{\omega}^{\eta}=\mathrm{diag}\big(\omega_{+}^{\eta},\omega_{-}^{\eta},-\omega_{+}^{\eta},-\omega_{-}^{\eta}\big)$ denotes a diagonal matrix. In particular,
its first component according to Eq.~\eqref{Eq:HartreeDensityResponseFunction}
gives the density response function
\begin{align*}
   \chi_{\mathrm{H}_{\eta};ij}^{R}(t)
   &=-(-1)^{i-j}\theta(t)\big(\chi_{+}^{\eta}\sin(\omega_{+}^{\eta}t)
   +\chi_{-}^{\eta}\sin(\omega_{-}^{\eta}t)\big)
   \, ,\\
   \chi_{\pm}^{\eta}
   &\equiv\frac{2t_{\mathrm{kin}}\big(\omega_{0}^{2}-\omega_{\pm}^{\eta}{}^{2}\big)^{2}}
   {\omega_{\pm}^{\eta}\Big[\big(\omega_{0}^{2}-\omega_{\pm}^{\eta}{}^{2}\big)^{2}
   +4\lambda \omega_{0}^{2}t_{\mathrm{kin}}^{2}\Big]}
   \, ,
\end{align*}
where we introduced the Heaviside function to enforce the correct causal
structure.

\newpage

\end{document}